\documentclass[english]{IEEEtran}
\usepackage[T1]{fontenc}
\usepackage{xcolor}
\usepackage{amsthm}

\usepackage{pdfcolmk}
\usepackage{float}
\usepackage{amsmath}
\usepackage{amsthm}
\usepackage{amssymb}
\usepackage{graphicx}
\usepackage{setspace}
\usepackage{wasysym}
\PassOptionsToPackage{normalem}{ulem}
\usepackage{ulem}
\usepackage{cite}
\makeatletter

\providecommand{\tabularnewline}{\\}
\floatstyle{ruled}
\newfloat{algorithm}{tbp}{loa}
\providecommand{\algorithmname}{Algorithm}
\floatname{algorithm}{\protect\algorithmname}
\providecolor{lyxadded}{rgb}{0,0,1}
\providecolor{lyxdeleted}{rgb}{1,0,0}

\DeclareRobustCommand{\lyxsout}[1]{\ifx\\#1\else\sout{#1}\fi}

\theoremstyle{plain}
\newtheorem{thm}{\protect\theoremname}
\theoremstyle{remark}
\newtheorem{rem}[thm]{\protect\remarkname}

\usepackage[caption=false, font=footnotesize]{subfig}

\usepackage{algorithmic}
\usepackage{cite}

\@ifundefined{showcaptionsetup}{}{%
	\PassOptionsToPackage{caption=false}{subfig}}
\usepackage{subfig}
\makeatother

\usepackage{babel}
\providecommand{\remarkname}{Remark}
\providecommand{\theoremname}{Theorem}

\begin{document}
\title{A Hybrid Reinforcement Learning Framework for Hard Latency Constrained
Resource Scheduling}
\author{{\normalsize Luyuan Zhang, An Liu, }{\normalsize\textit{Senior Member,
IEEE }}{\normalsize and Kexuan Wang\thanks{This work was supported in part by National Key Research and Development
Program of China under Grant 2021YFA1003304, in part by National Natural
Science Foundation of China under Grant 62071416, and in part by Zhejiang
Provincial Key Laboratory of Information Processing, Communication
and Networking (IPCAN), Hangzhou 310027, China.}\thanks{An Liu and Luyuan Zhang and Kexuan Wang are with the College of Information
Science and Electronic Engineering, Zhejiang University, Hangzhou
310027, China (email: \{22131107, anliu\}@zju.edu.cn).}}}
\maketitle
\begin{abstract}
In the forthcoming 6G era, extend reality (XR) has been regarded as
an emerging application for ultra-reliable and low latency communications
(URLLC) with new traffic characteristics and more stringent requirements.
In addition to the quasi-periodical traffic in XR, burst traffic with
both large frame size and random arrivals in some real world low latency
communication scenarios has become the leading cause of network congestion
or even collapse, and there still lacks an efficient algorithm for
the resource scheduling problem under burst traffic with hard latency
constraints. We propose a novel hybrid reinforcement learning framework
for resource scheduling with hard latency constraints (HRL-RSHLC),
which reuses polices from both old policies learned under other similar
environments and domain-knowledge-based (DK) policies constructed
using expert knowledge to improve the performance. The joint optimization
of the policy reuse probabilities and new policy is formulated as
an Markov Decision Problem\textcolor{black}{{} (MDP),} which maximizes
the hard-latency constrained effective throughput (HLC-ET) of users.
We prove that the proposed HRL-RSHLC can converge to KKT points with
an arbitrary initial point. Simulations show that HRL-RSHLC can achieve
superior performance with faster convergence speed compared to baseline
algorithms.
\end{abstract}


\section{Introduction\label{sec:introduction}}

\subsection{Background}

Ultra-reliable and low latency communications (URLLC), which is one
of the major communication scenarios of the fifth generation (5G)
wireless communication networks, has always been a key requirement
for many applications such as public safety, telemedicine and etc
\cite{you2021towards}\cite{saad2019vision}\cite{letaief2019roadmap}.
Average latencies are not of interest for URLLC applications, as an
instantaneous interruption in the transmitted data will lead to a
poor performance of the overall system. Meeting hard delay constraints
plays a critical role in supporting URLLC and existing works have
proposed various methods to meet the delay constraints. For example,
a novel framework utilizing a risk-constraint deep reinforcement learning
(DRL) algorithm was proposed in \cite{sulaiman2023generalizable}
to improve the scheduling performance by limiting the expectation
of quality of service (QoS) to a threshold. The author in \cite{huang20205g}
proposed a spectrum resource scheduling strategy based on the reinforcement
learning (RL) approach to meet the low latency constraints.

However, previous research efforts on the conventional URLLC use cases
in 5G only support short packets transmission, which cannot meet all
requirements of the future wireless communications. Researchers start
to focus on the sixth generation (6G) wireless communication networks,
and extend reality (XR) , which is an umbrella term for different
types of realities such as virtual reality (VR), augmented reality
(AR), and mixed reality (MR), has been regarded as an emerging application
for URLLC in 6G with new traffic characteristics and more stringent
requirements. Different from short-packet transmissions in conventional
URLLC, XR frame has a much larger size and requires multiple timeslots
to complete the transmission, which makes resource scheduling more
difficult to meet the hard delay constraints \cite{elbamby2018toward}.
To overcome this challenge, the author in \cite{chaccour2019reliability}
analytically derived the end-to-end delay distribution over a Terahertz
(THz) link, taking into account the VR frame's processing and transmission
delays. The author in \cite{chen2021frame} considered a cloud XR
architecture, formulated a resource allocation problem to maximize
the number of satisfied users under the data rate, reliability and
latency requirements for each user, and solved the problem approximately
by jointly considering the user admission control and a frame-level
integrated transmission problem.

In addition to the quasi-periodical traffic in XR, burst traffic with
both large frame size and random arrivals in some real world low latency
communication scenarios has become the leading cause of network congestion
or even collapse \cite{yu2020burst}. However, to the best of our
knowledge, existing works in the literature have not considered resource
scheduling with hard delay constraints under burst traffic, which
still face various technical challenges to be addressed.

\subsection{Related Works}

There are two major methods to solve resource scheduling problems
with delay constraints.

\subsubsection{Non-RL methods}

One common approach to deal with scheduling problems with latency
constraints is to assume simple and known traffic/channel statistics.
The author in \cite{kharel2022fog} assumed the distribution of the
traffic and channel was known, then attained accurate closed-form
expressions for the outage probability, and proposed a mini-slots-based
scheduling framework to serve URLLC users under latency deadlines.
A hybrid multiple access (HMA) solution was proposed in \cite{azari2019risk}
based on the outage probability for URLLC traffic. In \cite{hassan2010energy},
it was assumed that all the packets are of equal size and have a strict
delay constraint, and the objective of the scheduler is to minimize
the total transmit power subject to strict delay constraints and the
author developed the upper and lower bound on output rates and developed
a two step solution. \cite{shan2014energy} considered a multi user
multiple-input multiple-output (MU-MIMO) with delay constraints, in
which a certain size of data in each user's queue needs to be transmitted
within the deadline. The author in \cite{hassan2012resource} considered
a downlink OFDMA system with at most one packet in each user's queue,
in which all the packets cannot be delivered before their deadline
would be dropped, formulated a constrained resource optimization problem
and then replaced the reverse convex constraint by a linear constraint,
transforming the original problem into a convex optimization problem,
and achieved the delay constraints for each packet. \cite{hassan2010energy},
\cite{shan2014energy} and \cite{hassan2012resource} all assume known
traffic/channel statistics, which is difficult to achieve in real
world application scenarios with burst traffic and dynamic wireless
environment.

\subsubsection{RL methods}

RL method is a common approach to get rid of the unrealistic assumptions
on traffic/channel statistics. The author in \cite{kasgari2019model}
proposed a deep-RL framework to jointly allocate resource blocks (RBs)
and power, and dynamically measures the end to end reliability and
the delay of each user, achieving the low latency constraints. The
author in \cite{saggese2021deep} proposed a DRL approach to train
an agent acting as a scheduler and never violate URLLC latency requirements.
\cite{fawaz2018optimal} formulated the scheduling problem with a
strict delay constraint on each queued packet stored in the buffer
as a Markov Decision Problem (MDP) and solved it using relative value
iteration (VI) algorithm. To obtain the optimal trade-off between
delay and power consumption for a given power constraint in a communication
system whose traffic/channel conditions can change over time, the
author in \cite{zhao2019reinforcement} formulated the problem as
an infinite-horizon MDP and then Q-learning was adopted to solve this
problem. The author in \cite{shafieirad2022meeting} proposed the
use of RL and deep learning to address the max delay latency constraint
through a sequential decision making model. However, the above existing
works on RL-based resource scheduling focus on conventional URLLC
or other simple scenarios, and have not considered both hard delay
constraints and burst traffic.

\subsection{Contributions}

The above existing works in the literature have not fully addressed
the problem of resource scheduling with hard delay constraints under
burst traffic. Although there have been some attempts to apply RL
for this problem, the large state/action space, the hard latency constraints
and the more unpredictable burst traffic with multi-timeslots transmissions
make the conventional RL algorithms hard to converge. To solve this
challenging problem, we propose a novel resource scheduling algorithm
called HRL-RSHLC in this paper. The main contributions of this work
are:
\begin{itemize}
\item \textbf{A hybrid RL framework for burst traffic with hard latency
constraints: }Unlike the conventional RL algorithms which aim to optimize
a single policy, we propose a hybrid RL framework consists of a mixture
of a deep neural network (DNN) parameterized policy (which is called
the new policy), a domain-knowledge-based policy constructed using
expert knowledge (which is called the DK policy) and old policies
trained under other similar environments. The joint optimization of
the probabilities for using each policy and DNN parameters of the
new policy are formulated as an \textcolor{black}{MDP, }where the
objective is to maximize the hard-latency constrained effective throughput
(HLC-ET) of users. In particular, the hard delay constraints are embodied
in the objective function, that is, only packets which have been successfully
delivered before the hard delay constraints would be considered in
the HLC-ET, which avoid the use of constrained MDP (CMDP). However,
due to the unpredictable burst traffic with multi-timeslots transmissions,
the implicit hard delay constraints in the objective function also
brings sparse rewards, which poses challenges for fast-convergent
algorithm design.
\item \textbf{A fast convergent resource scheduling algorithm based on hybrid
RL to maximize the HLC-ET:} Based on the above hybrid RL framework,
we propose a novel resource scheduling algorithm called HRL-RSHLC,
to solve the MDP of maximizing the HLC-ET. In HRL-RSHLC, both policy/data
reuse as well as the domain specific knowledge are exploited to accelerate
the convergence speed. Specifically, instead of directly controlling
the user scheduling and MIMO precoder, the HRL-RSHLC only controls
a priority weight vector, and the user scheduling and MIMO precoder
are indirectly determined by maximizing the weighted sum-rate (WSR)
using a classic iterative algorithm. Such a design can significantly
reduce the action space. The optimization of policy reuse probabilities
can further accelerate the convergence speed. If an old policy performs
well in the current environment, its reuse probability will automatically
be increased by the hybrid RL algorithm to accelerate the initial
convergence speed. Moreover, we introduce a DK policy in which the
action (weight vector) is chosen as Q-weighted greedy scheduling algorithm
\cite{dimic2005downlink}. The DK policy is shown to perform well
for light to moderate traffic loading, and can provide a stable scheduling
performance. As such, even if all the DNN-based policies cannot work
well (e.g., due to the limited representational capacity of practical
DNNs and the lack of interpretability), the hybrid RL algorithm will
automatically increase the reuse probability of the DK policy to avoid
any catastrophic failure of DNN-based policies. Finally, HRL-RSHLC
updates the policy using both stored old experiences and newly added
data, and alleviates the issues caused by the sparsity of reward.
\item \textbf{Convergence analysis and theoretical performance guarantee:}
We prove that the proposed HRL-RSHLC algorithm can converge to KKT
points with an arbitrary initial point. Simulations show that HRL-RSHLC
can achieve superior performance with faster convergence speed compared
to baseline algorithms.
\end{itemize}
The rest of the paper is organized as follows. In Section \ref{sec:System-model},
we illustrate the system model considered in this paper. In Section
\ref{sec:framework}, we elaborate the hybrid RL framework and problem
formulation for burst traffic with hard latency constraints. In Section
\ref{sec:algorithm}, we propose a HRL-RSHLC algorithm to solve the
formulated resource scheduling problem. In Section \ref{sec:converge},
we prove that the proposed HRL-RSHLC can converge to KKT points with
an arbitrary initial point. Section \ref{sec:Simulation-Results}
showcases the simulation results. Finally, we conclude this paper
in Section \ref{sec:conclusion}.

\section{System Model \label{sec:System-model}}
\begin{figure*}[tp]
	\begin{centering}
		\textsf{\protect\includegraphics[scale=0.55]{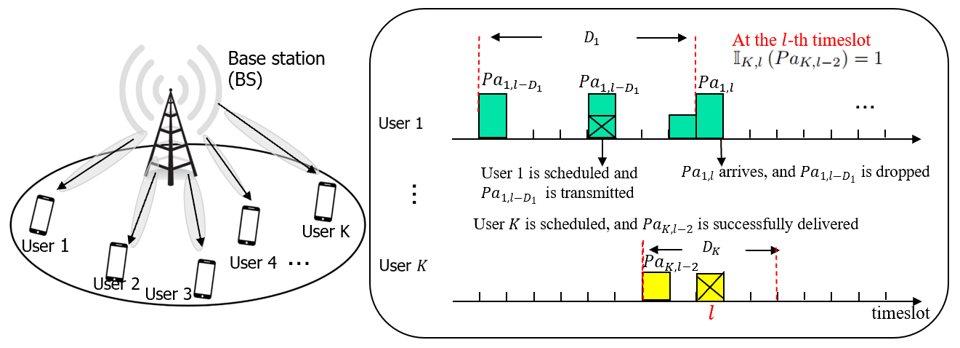}}\protect
		\par\end{centering}
	\caption{\label{fig:system-model-1}System model}
\end{figure*}

We consider a downlink multi-user MIMO (MU-MIMO) system under burst
traffic. As shown in Fig. 1, a base station (BS) serves a set of $K$
users $\mathcal{K}$. The BS is equipped with $N_{T}$ antennas and
each user is equipped with a single antenna. At the BS, each user
is assigned an individual buffer. Data packets from some higher-layer
applications randomly arrive at the buffers and are stored in the
form of queues and the BS dynamically schedules the resource for the
downlink transmission of each data queue.

\subsection{Downlink Multi-user MIMO (MU-MIMO) Transmission}

The proposed algorithm is efficient for any precoding scheme and power
allocation scheme. In the simulations, we adopt regularized zero-forcing
(RZF) and equal power allocation under the single-stream transmission
mode, which are widely applied in real systems.

We suppose that time is divided into time slots of duration $\tau$
with index $l$. Let $pow_{i}$ and $\boldsymbol{v}_{i}\in\mathbb{C}^{N_{T}\times1}$
denote the transmit power and normalized precoder for user $i$. At
the $l$-th time slot, the received signal at user $i$ can be formulated
as:
\begin{equation}
\boldsymbol{y}_{i}=\boldsymbol{h}_{i}\sqrt{pow_{i}}\boldsymbol{v}_{i}d_{i}+\sum_{j\in\mathcal{B}/i}\boldsymbol{h}_{i}\sqrt{pow_{j}}\boldsymbol{v}_{j}d_{j}+\boldsymbol{n}_{i},
\end{equation}
where $\boldsymbol{h}_{i}\in\mathbb{C}^{1\times N_{T}}$ is the channel
vector of user $i$, the time slot index $l$ is omitted for conciseness,
and $\mathcal{B}$ denotes the scheduled user set, which will be given
in detail in Section \ref{subsec:Resource-Scheduling-Based}. $d_{i}$
is the normalized data symbols, i.e., $\mathbb{E}\left[|d_{i}|^{2}\right]=1$,
and $\boldsymbol{n}_{i}\sim\mathcal{CN}\left(0,\sigma_{i}^{2}\right)$
is Additive White Gaussian Noise (AWGN) with variance $\sigma_{i}^{2}$.
Therefore, the signal-to-interference-plus-noise-ratio (SINR) of user
$i$ can be formulated as
\begin{equation}
\textrm{SINR}_{i}=\frac{pow_{i}|\boldsymbol{h}_{i}^{H}\boldsymbol{v}_{i}|^{2}}{\sum_{j\in\mathcal{B}/i}pow_{j}|\boldsymbol{h}_{i}^{H}\boldsymbol{v}_{j}|^{2}+\sigma_{i}^{2}},
\end{equation}
then the data rate of $i$-th user is given by
\begin{equation}
R_{i}=B\log_{2}\left(1+\frac{pow_{i}|\boldsymbol{h}_{i}^{H}\boldsymbol{v}_{i}|^{2}}{\sum_{j\in\mathcal{B}/i}pow_{j}|\boldsymbol{h}_{i}^{H}\boldsymbol{v}_{j}|^{2}+\sigma_{i}^{2}}\right),\forall i
\end{equation}
where $B$ is the bandwidth, and $\boldsymbol{V}=\left[\boldsymbol{v}_{i}\right]{}_{i\in\mathcal{K}}\in\mathbb{C}^{N_{T}\times|\mathcal{B}|}$
is the normalized RZF precoding matrix, which is given by
\begin{equation}
\boldsymbol{V}=\boldsymbol{H}_{B}^{H}\left(\boldsymbol{H}_{B}\boldsymbol{\boldsymbol{H}}_{B}^{H}+\alpha\boldsymbol{I}\right)^{-1}\boldsymbol{\lambda}^{\frac{1}{2}},\label{eq:precoder}
\end{equation}
where $\alpha$ is a regularization factor, $\boldsymbol{\lambda}$
is a diagonal matrix for normalization, and $\boldsymbol{H}_{B}\left(t\right)\in\mathbb{C}^{|\mathcal{B}|\times N_{T}}$
is channel matrix formed by merging the equivalent channels of scheduled
users.
\begin{rem}
For clarity, we assume that the users have a single-antenna and the
BS has perfect channel state information (CSI). As will be discussed
in Section \ref{subsec:Resource-Scheduling-Based}, the proposed resource
allocation algorithm does not directly control the MIMO precoder but
only control the priority weight of each user. As such, the proposed
algorithm does not rely on any specific MU-MIMO transmission scheme
at the physical layer, and it can be directly applied in a system
when the users have multiple antennas or the BS has imperfect CSI.
For example, when the BS has imperfect CSI, the precoder can be calculated
using the the estimated channel, i.e., in (\ref{eq:precoder}), $\boldsymbol{H}_{B}$
is replaced by the estimated channel $\hat{\boldsymbol{H}_{B}}$.
If the users have multiple antennas, we can simply use the block diagonal
(BD) precoder or any other precoder designed for the case with multi-antenna
users.
\end{rem}
The key notations are summarized in Table \ref{tab:key notations}.

\begin{table}[t]
\centering{}\caption{\textcolor{blue}{\label{tab:key notations}}Summary of Notations}
\begin{tabular}{|c|c|}
\hline 
{\scriptsize$K$/$\mathcal{K}$} & {\scriptsize Number of users/ User set}\tabularnewline
\hline 
{\scriptsize$N_{T}$/$N_{R}$} & {\scriptsize Number of BS/user antenna}\tabularnewline
\hline 
{\scriptsize$\boldsymbol{h}_{i}$} & {\scriptsize Channel of user $i$}\tabularnewline
\hline 
{\scriptsize$l$} & {\scriptsize Time slot index}\tabularnewline
\hline 
{\scriptsize$pow_{i}$/$\boldsymbol{v}_{i}$} & {\scriptsize Power/precoder of user $i$}\tabularnewline
\hline 
{\scriptsize$Pa_{i,l}$} & {\scriptsize Packet that arrives at user $i$}\tabularnewline
\hline 
{\scriptsize$\bar{Q}_{i,l}$} & {\scriptsize Original length of $Pa_{i,l}$}\tabularnewline
\hline 
{\scriptsize$Q_{i,l}$} & {\scriptsize Remaining length of $Pa_{i,l}$}\tabularnewline
\hline 
{\scriptsize$PA$} & {\scriptsize Packet arrival probability}\tabularnewline
\hline 
{\scriptsize$D_{i}$} & {\scriptsize Delay constraint for user $i$}\tabularnewline
\hline 
{\scriptsize$R_{i}$} & {\scriptsize Data rate of user $i$}\tabularnewline
\hline 
{\scriptsize$p_{n}$} & {\scriptsize Probability to use policy $n$}\tabularnewline
\hline 
{\scriptsize$\tau$} & {\scriptsize Time slot duration}\tabularnewline
\hline 
\end{tabular}
\end{table}

\subsection{Traffic and Queue Dynamic Model}

We assume that the data packets only arrive at the start of each timeslot.
Specifically, at the $l$-th timeslot, a single data packet $Pa_{i,l}$
of length $\bar{Q}_{i,l}$ Kbit arrives at the queue of user $i$
with a probability $PA$. The length of arrived data is random with
$\mathbb{E}\left(\bar{Q}_{i,l}\right)=\lambda_{i}$.

The delay constraint for user $i$ is $D_{i}$ timeslots,\textcolor{blue}{{}
}which means that if a packet arrives at user $i$'s queue at the
$l$-th timeslot, and at the $(l+D_{i})$-th timeslot it has not been
successfully delivered, then it would be dropped out of the queue
at this timeslot. Apparently, there are at most $D_{i}$ packets in
the queue of user $i$. To better capture the state of each packet
in the queue, we denote $Q_{i,l}$ as the remaining data size of the
packet $P_{i,l}$, and $B\left(Pa_{i,l}\right)=\sum_{l'=1}^{D_{i}-1}Q_{i,l-l'}$
as the length of the packet backlog in front of $Pa_{i,l}$. The arrived
data packets are served according to the first-come-first-served (FCFS)
protocol. In the FCFS protocol, packets are serviced in the order
they arrive, that is, the first packet to arrive is the first one
to be transmitted, followed by the next one in queue, and so on. $Q_{i,l}$
is the remaining size of packet $Pa_{i,l}$ and $B\left(Pa_{i,l}\right)=\sum_{l'=1}^{D_{i}-1}Q_{i,l-l'}$
is the length of the packet backlog in front of $Pa_{i,l}$. Thus,
according to the FCFS protocol, the packet $Pa_{i,l}$ will not be
served until $B\left(Pa_{i,l}\right)=0$. In hard-latency constrained
transmissions, we focus on the following two crucial cases:
\begin{itemize}
\item \textbf{Data being dropped:} Packets failed to be delivered before
their deadlines would be dropped. Specifically, at the $l$-th time
slot, the packet $Pa_{i,l-D_{i}}$in the queue of user $i$ would
be dropped if $Q_{i,l-D_{i}}>R_{i}\left(l-1\right)\tau$.
\item \textbf{Data being successfully delivered:} We define a binary functions
$\mathbb{I}_{i,l}\left(Pa_{i,l'}\right)$, $\forall i,l$, to indicate
whether the packet $Pa_{i,l'}$ is successfully delivered at the $l$-th
time slot:
\begin{equation}
\mathbb{I}_{i,l}\left(Pa_{i,l'}\right)=\begin{cases}
1, & \mathrm{if}\ B\left(Pa_{i,l'}\right)+Q_{i,l'}\leq R_{i}\left(l\right)\tau,\\
0, & \mathrm{otherwise}.
\end{cases}
\end{equation}
\end{itemize}
For ease of understanding, we show a possible state of the data queues
in Fig. \ref{fig:system-model-1}. As shown in Fig. \ref{fig:system-model-1},
packet $Pa_{1,l-D_{1}}$ arrived at user $1$'s queue $D_{1}$ timeslots
ago, then user 1 was scheduled 4 timeslots ago and part of the packet
was transmitted, which is the green section marked with an X in the
figure. At the $l$-th timeslot, packet $Pa_{1,l-D_{1}}$ has not
been successfully delivered ($R_{1}(l-1)\tau<$$Q_{1,l-D_{1}}$) so
it is \textit{dropped} and a new packet $Pa_{1,l}$ arrives at user
1's queue at the same time. At the $l$-th timeslot, user $K$ is
scheduled and its transmission rate $R_{K}(l)\tau\geq Q_{K,l-2}$,
so $Pa_{K,l-2}$ is \textit{successfully delivered}.

\subsection{Hard-Latency Constrained Effective Throughput\label{subsec:Hard-Latency-Constrained-Effecti}}

In this paper, we focus on optimizing the hard-latency constrained
effective throughput (HLC-ET) of users. Specifically, at the $l$-th
time slot, the instantaneous HLC-ET of users can be defined as
\begin{equation}
\frac{1}{\tau}\sum_{i=1}^{K}\sum_{l'=0}^{D_{i}-1}\mathbb{I}_{i,l}\left(Pa_{i,l-l'}\right)\bar{Q}_{i,l-l'}.\label{eq:HLC-ET}
\end{equation}
where $\tau$ is the duration of each timeslot, $Pa_{i,l-l'}$ is
the packet that arrived at user $i$'s queue at $l-l'$-th timeslot,
and $\overline{Q}_{i,l-l'}$ is the original length of packet $Pa_{i,l-l'}$.
The delay constraint of user $i$ is $D_{i}$, so at the $l$-th timeslot,
the oldest packet in user $i$'s queue is $Pa_{i,l-(D_{i}-1)}$. The
indication function $\mathbb{I}_{i,l}\left(Pa_{i,l-l'}\right)=1$
means that $Pa_{i,l-l'}$ with the original packet length $\bar{Q}_{i,l-l'}$
is successfully delivered at the $l$-th timeslot, which contributes
to the HLC-ET with the term $\frac{1}{\tau}\bar{Q}_{i,l-l'}$.\textcolor{blue}{{}
}Without loss of generality, we set $\tau=1$. When packet $Pa_{i,l-l'}$
is delivered successfully at the $l$-th time slot, the original data
size $\bar{Q}_{i,l-l'}$ would be included in the HLC-ET. To simplify
the presentation, we define vectors $\boldsymbol{Q}\left(l\right)=\left[Q_{1,l-D_{1}+1},...,Q_{1,l},...,Q_{K,l-D_{K}+1},...,Q_{K,l}\right]^{T}\in\mathbb{R}^{\sum_{i=1}^{K}D_{i}}$
and $\bar{\boldsymbol{Q}}\left(l\right)=\left[\bar{Q}_{1,l-D_{1}+1},...,\bar{Q}_{1,l},...,\bar{Q}_{K,l-D_{K}+1},...,\bar{Q}_{K,l}\right]^{T}\in\mathbb{R}^{\sum_{i=1}^{K}D_{i}}$.

\section{Hybrid RL Framework and Problem Formulation\label{sec:framework}}

In this section, we propose a hybrid RL framework to achieve efficiently
online scheduling in the considered system. First, we explain the
motivation of choosing the priority weight as the action of the agent.
Then, the problem formulation based on MDP is presented. Finally,
we introduce the details of the hybrid RL framework.

\subsection{Resource Scheduling Based on Weighted Sum-Rate Maximization\label{subsec:Resource-Scheduling-Based}}

Some previous works applying the RL algorithms to solve resource scheduling
problems have chosen discrete actions, e.g., the scheduling algorithm
proposed in \cite{fawaz2018optimal} chose the number of transmitted
packets as the action. However, since the values of the states, e.g.,
the CSI, are continuous, applying discrete actions may decrease the
performance of RL. On the other hand, the action space of discrete
actions would be large when MU scheduling is considered, extremely
when the number of users is large.

\begin{figure}[tp]
\protect\begin{centering}
\textsf{\protect\includegraphics[width=6.5cm,height=4.7cm]{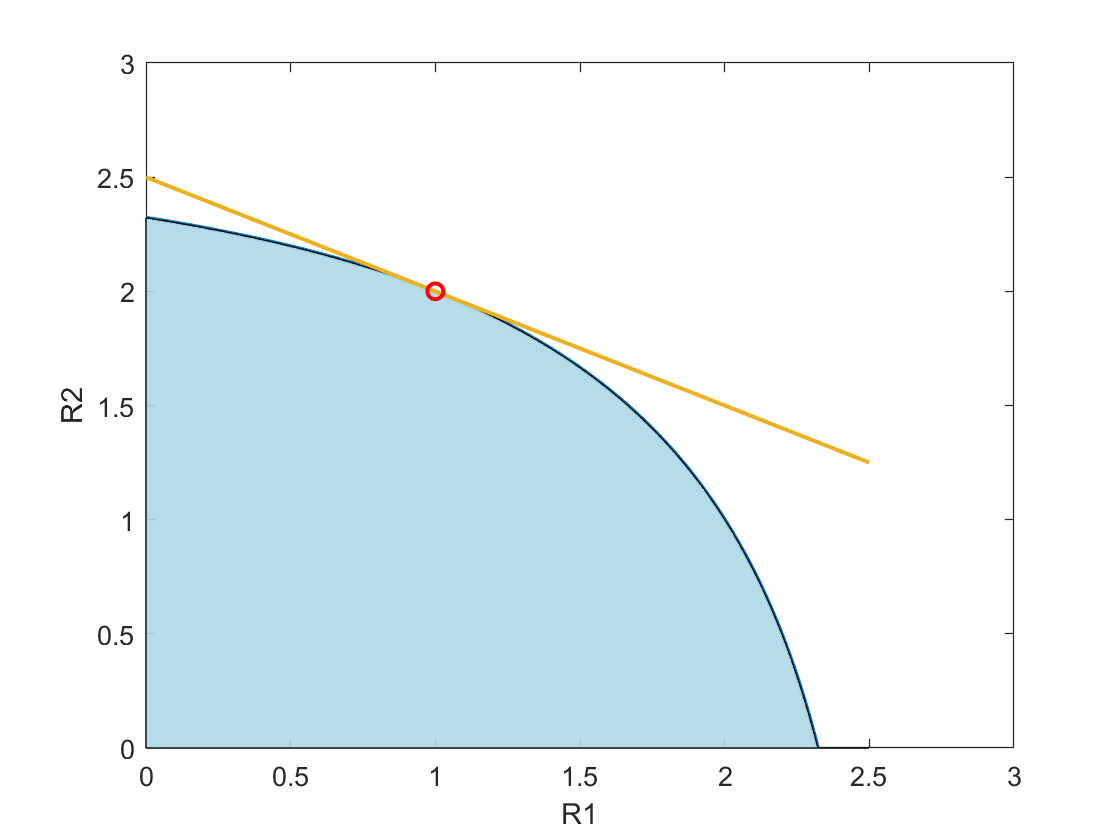}}\protect
\par\end{centering}
\protect\caption{\label{fig:rate region}Illustration of a strongly convex rate region.}
\end{figure}

In this paper, we design a resource scheduling algorithm based on
controlling the priority weights of users, and the user scheduling
and MIMO precoder/power allocation are indirectly determined by maximizing
the weighted sum-rate (WSR) using a classic iterative algorithm. Such
a design can significantly reduce the action space and speed up the
convergence compared to directly controlling all of the variables,
i.e., the user scheduling and power allocation. When the data rate
region is strongly convex, using the priority weights as the control
action and maximizing the WSR will not lose any optimality, as explained
below. For a given weight, the unique WSR maximization rate point
is the tangent point of the plane determined by the weight and the
rate region, as shown in Fig. \ref{fig:rate region}. This means that
any Pareto rate point on the boundary of the strongly convex rate
region can be achieved by maximizing the WSR with a proper weight
vector. Since the optimal user scheduling and power allocation must
also achieve a certain Pareto rate point, and we can always achieve
the same Pareto rate point by controlling the weight vector, directly
controlling the weight vector will not lose any optimality.

It is well-known that the capacity region of Gaussian MIMO broadcast
channel (BC) is strongly convex under a total power constraint. Thus
directly controlling the priority weights will not lose any optimality
for capacity achieving physical layer schemes (such as dirty paper
coding \cite{weingarten2006capacity}). Simulations show that directly
controlling the priority weights is still very efficient for resource
allocation under other sub-optimal but more practical physical layer
scheme such as RZF beamforming \cite{dimic2005downlink}, even when
the rate region is not strongly convex in this case. Indeed, we present
simulation results in Section \ref{sec:Simulation-Results} that compares
the performance of choosing the weight as action and directly controlling
the user scheduling and power allocation in the action. Simulation
results show that reducing the action space can significantly improve
the convergence speed. In fact, HRL-RSHLC controlling all actions
failed to converge to a good solution, which is probably due to fact
that an enlarged action space makes it more likely to get stuck in
a bad local optimum.

We define the priority weight vector as $\boldsymbol{w}=\left[w_{1},...,w_{K}\right]$,
then in the proposed design, the optimal user scheduling and power
allocation scheme is obtained by maximizing the WSR of users. The
optimization problem can be formulated as:
\begin{equation}
\max_{\mathcal{B},\left\{ pow_{i}\right\} _{i\in\mathcal{\mathcal{B}}}}\sum_{i\in\mathcal{B}}w_{i}\cdot B\log_{2}\left(1+\frac{pow_{i}|\boldsymbol{h}_{i}^{H}\boldsymbol{v}_{i}|^{2}}{\sum_{j\in\mathcal{B}/i}pow_{j}|\boldsymbol{h}_{i}^{H}\boldsymbol{v}_{j}|^{2}+\sigma_{i}^{2}}\right),\label{WSR-1}
\end{equation}
where $\mathcal{B}$ denotes the scheduled user set, $\left\{ pow_{i}\right\} _{i\in\mathcal{\mathcal{B}}}$
is determined according to equal power allocation scheme, and $pow_{i}=0,\forall i\notin\mathcal{\mathcal{B}}$.
Note that the proposed hybrid RL framework works for any resource
allocation policy that aims at maximizing the WSR, and we adopt the
greedy user selection algorithm with RZF MIMO precoder and equal power
allocation \cite{dimic2005downlink} in this paper due to its wide
application in practical systems. We assume that the BS knows the
channel state information (CSI) of all users and the greedy user selection
algorithm with RZF MIMO precoder is based on the known CSI at the
BS. Specifically, the greedy user scheduling algorithm selects users
by round: In each round, it finds one user that be added to the selected
user set to maximize the WSR, until no more users can be found that
would increase the WSR. Clearly, the proposed RL framework based on
controlling priority weights works for any other MU-MIMO transmission
scheme, as long as one can design an iterative algorithm to solve
the WSR maximization problem.

\subsection{Problem Formulation based on MDP}

An MDP is denoted as a tuple $\left(\mathcal{S},\mathcal{A},R,P\right),$
where $\mathcal{S}$ is the state space, $\mathcal{A}$ is the action
space, $R:\mathcal{S}\times\mathcal{A}\rightarrow\mathbb{R}$ is the
reward function. $P:\mathcal{S}\times\mathcal{A\times\mathcal{S}}\rightarrow\left[0,1\right]$
is the transition probability function, and $P\left(\boldsymbol{s}'\mid\boldsymbol{s},\boldsymbol{a}\right)$
denotes the transition probability from state $s$ to state $s'$
under action $\boldsymbol{a}$. A policy $\pi:\mathcal{S}\rightarrow\mathbf{P}\left(\mathcal{A}\right)$
is a map from states to probability distributions over actions, and
$\pi\left(\boldsymbol{a}\mid\boldsymbol{s}\right)$ denotes the probability
of choosing action $\boldsymbol{a}$ in state $\boldsymbol{s}$. Due
to the curse of dimensionality, modern RL algorithms, e.g., deep reinforcement
learning (DRL)-based algorithms, usually parameterize the policy by
function approximations with high representation capability, e.g.,
DNN. In this paper, we denote $\pi_{\boldsymbol{\theta}}$ as the
policy parameterized by $\boldsymbol{\theta}$, as will be detailed
later in Subsection \ref{subsec:Hybrid-RL-Framework}.
\begin{itemize}
\item \textbf{The state space} $\mathcal{S}$\textbf{: }$\mathcal{S}$ is
a composite space consisting of the queue state space and the channel
state space, i.e., the current state information at the $l$-th time
slot is denoted as $\boldsymbol{s}_{l}=\left\{ \boldsymbol{Q}\left(l\right),\bar{\boldsymbol{Q}}\left(l\right),\boldsymbol{H}\left(l\right)\right\} $,
where $\boldsymbol{H}\left(t\right)\in\mathbb{C}^{K\times N_{T}}$
is channel matrix formed by merging the equivalent channels of all
users.
\item \textbf{The action space} $\mathcal{A}$\textbf{:} the priority weight
vector space constitute the action space $\mathcal{A}$, i.e., the
action at the $l$-th time slot is $\boldsymbol{a}_{l}=\bigl\{\boldsymbol{w}_{l}\bigr\}$.
Specifically, the action $\boldsymbol{a}_{l}$ is sampled according
to a policy $\pi_{\boldsymbol{\theta}}:\mathcal{S}\rightarrow\mathbf{P}\left(\mathcal{A}\right)$.
\item \textbf{The transition probability function }$P$\textbf{:} the function
$P:\mathcal{S}\times\mathcal{A}\times\mathcal{S}\rightarrow\left[0,1\right]$
is an unknown transition probability function related to the statistics
of the unknown statistics of environment model, where $P\left(\boldsymbol{s}_{l+1}\mid\boldsymbol{s}_{l},\boldsymbol{a}_{l}\right)$
denotes the probability of transition to state $\boldsymbol{s}_{l+1}$
from state $\boldsymbol{s}_{l}\in\mathcal{S}$ with an action $\boldsymbol{a}_{l}$.
The transition probability $P$ and policy $\pi_{\boldsymbol{\theta}}$
together determine the probability distribution of the trajectory
$\left\{ \boldsymbol{s}_{0},\boldsymbol{a}_{0},\boldsymbol{s}_{1},\ldots\right\} $.
\item \textbf{The reward function} $R$: at each timeslot $l$, the instantaneous
HLC-ET of users is set to be the reward. The greedy user scheduling
algorithm is applied to maximize the WSR based on $\boldsymbol{a}_{l}$,
and we have $R\left(\boldsymbol{s}_{l},\boldsymbol{a}_{l}\right)=\sum_{i=1}^{K}\sum_{l'=0}^{D_{i}-1}\mathbb{I}_{i,l}\left(Pa_{i,l-l'}\right)\bar{Q}_{i,l-l'}$.
\end{itemize}
In this paper, we consider the problem of maximizing the average HLC-ET,
which can be formulated as an MDP:
\begin{align}
\min_{\boldsymbol{\theta}\in\mathsf{\Theta}}J\left(\theta\right) & \triangleq\lim_{L\rightarrow\infty}\frac{1}{L}\mathbb{E}_{p_{s}\sim\pi_{\boldsymbol{\theta}}}[-\sum_{l=0}^{L-1}\sum_{i=1}^{K}\sum_{l'=0}^{D_{i}-1}\mathbb{I}_{i,l}\left(Pa_{i,l-l'}\right)\cdot\nonumber \\
 & \bar{Q}_{i,l-l'}],\label{eq:objective}
\end{align}
where $p_{s}\sim\pi_{\boldsymbol{\theta}}$ denote the probability
distribution of the trajectory under policy $\pi_{\boldsymbol{\theta}}$.
Note that there is no need to add an explicit constraint for the probability
of violating the hard delay constraint due to the following reason.
When all of the packets have the same size and delay constraint $D_{max}$,
the average HLC-ET is equal to the product of the packet arrival rate
and the successful transmission probability, i.e. $A*\left(1-\Pr\left(D>D_{max}\right)\right)$,
where $A$ is the packet arrival rate. Therefore, maximizing the HLC-ET
is equivalent to minimizing the probability of violating the hard
delay constraint $\Pr\left(D>D_{max}\right)$.

Another possible formulation is to maximize the average throughput
with an explicit constraint for the probability of violating the hard
delay constraint, i.e., considering the following constrained MDP
(CMDP):

\begin{equation}
\min_{\boldsymbol{\theta}\in\mathsf{\Theta}}J\left(\theta\right)\triangleq\lim_{L\rightarrow\infty}\frac{1}{L}\mathbb{E}_{p_{s}\sim\pi_{\boldsymbol{\theta}}}\left[-\sum_{l=0}^{L-1}\sum_{i=1}^{K}\textrm{THP}_{i}\left(l\right)\right]
\end{equation}
\[
\textrm{s.t. \ensuremath{\lim_{L\rightarrow\infty}\frac{1}{L}\mathbb{E}_{p_{s}\sim\pi_{\boldsymbol{\theta}}}\left[\sum_{l=0}^{L-1}\frac{u'\left(l\right)}{u\left(l\right)}\right]}}-\epsilon\leq0
\]
where $\textrm{THP}_{i}\left(l\right)=\min\left(R_{i}\left(l\right),\sum_{j=1}^{D_{i}}Q_{l-j+1}\right)$
represents the throughput of user $i$, $u'\left(l\right)$ is the
number of packets transmitted to users in timeslot $l$, whose latency
exceeds $D_{max}$, $u\left(l\right)$ is the total number transmitted
to users, and $\epsilon$ is the maximum allowable probability of
violating the hard delay constraint. $\frac{u'\left(l\right)}{u\left(l\right)}$
is the estimated value of $\left(1-\Pr\left(D>D_{max}\right)\right)$
at the $l$-th timeslot \cite{kasgari2019model}.

In Fig. \ref{fig:a-1} and \ref{fig:b-1}, we compare the probability
of ensuring hard latency for the HRL-RSHLC algorithms based on the
MDP and CMDP formulations respectively. It is observed from the simulations
that solving the CMDP problem requires higher computational complexity.
Moreover, the MDP-based HRL-RSHLC can achieve a higher effective throughput
because it is designed to directly maximize the effective throughput
and it also has a better convergence behavior. Thus, in this paper,
we shall design the HRL-RSHLC algorithm based on the MDP formulation
in (\ref{eq:objective}).

\subsection{Hybrid RL Framework\label{subsec:Hybrid-RL-Framework}}

To accelerate the convergence speed and reduce the interaction costs
with the environment in RL, various methods have been proposed in
the literature and most of them consider transferring learning (TL)
or domain knowledge exploitation, which are heuristic without rigorous
convergence analysis or theoretical performance guarantee, e.g., the
probability of using each policy is determined by a heuristic method
and the convergence of the overall algorithm is not guaranteed in
the policy reuse method \cite{fernandez2006probabilistic}. Since
the proposed hybrid RL framework is inspired by the policy reuse method,
we shall first review this method as preliminary and motivation for
the proposed framework.

\subsubsection{Preliminary on the Policy Reuse}

Policy reuse is a technique for reinforcement learning guided by past
learned similar policies. Policy reuse method relies on using the
past policies as a probabilistic bias where the learning agent faces
three choices: the exploitation of the ongoing learned policy, the
exploration of random unexplored actions, and the exploitation of
past policies. The key component of policy reuse is a similarity function
to estimate the similarity of past policies with respect to a new
one. Specifically, let $W_{i}$ be the gain obtained while reusing
the past policy $\pi_{i}$, and policy reuse method use such value
to measure the usefulness of reusing the policy $\pi_{i}$ to learn
the new policy $\pi_{new}$. Policy reuse introduces a solution consists
of following a softmax strategy using the values $W_{i}$ and $W_{new}$,
where the probability of using each policy can be expressed as
\begin{equation}
\Pr\left(\pi_{i}\right)=\frac{\exp\left(\nu W_{i}\right)}{\sum_{j=0}^{N}\exp\left(\nu W_{j}\right)}
\end{equation}
where $\nu$ is a temperature parameter, and $W_{new}=W_{0}$ is the
average reward when following the current learned policy $\pi_{new}$.
This provides a way to decide whether to exploit the past policies
or the new one.

Practice has shown that when a very similar policy is included in
the set of policies to be reused, the improvement on learning is very
high, and when the algorithm discovers that reusing the past policies
is not useful, it will follow the best strategy available, which is
the new policy. As such, policy reuse method in \cite{fernandez2006probabilistic}
can be viewed as a stochastic policy, whose actions are generated
from a mixture distribution (e.g., mixture Gaussian policy) instead
of a single distribution, e.g., the Gaussian policy \cite{sutton2018reinforcement}.

\subsubsection{Motivation for the Proposed Hybrid RL Framework}

In policy reuse method, the probability of using each sub-policy,
i.e., following each sub-distribution, is determined by a heuristic
method and the convergence of the overall algorithm is not guaranteed.
In other words, only the parameter of the new policy is optimized
based on RL. Thus, we propose a\textit{ hybrid RL framework}, which
essentially can also be viewed as a stochastic policy with parameters
$\boldsymbol{\theta}=\left[\boldsymbol{p};\boldsymbol{\gamma}_{0}\right]\in\boldsymbol{\Theta}$
to be optimized based on the policy gradient, where $\boldsymbol{\gamma}_{0}$
is the parameter of the new policy and $\boldsymbol{p}$ is the probability
of using each sub-policy. Such a stochastic policy can be seen as
a generalization of the conventional stochastic policy with only a
single sub-distribution, e.g., the Gaussian policy is a special case
when there is only one sub-policy/sub-distribution. In addition, the
hybrid RL framework contains not only old similar policies but also
the DK policy to improve robustness. This is because the DK policy
constructed using expert knowledge is more robust (i.e., has better
generalization ability and interpretability), but usually has some
gap w.r.t. the optimal solution, while old policies trained under
other similar environments can provide a good initial performance
to accelerate the convergence but sometimes are not robust (e.g.,
due to the limited representational capacity of practical DNNs and
the lack of interpretability). The proposed hybrid RL framework achieves
robustness as well as fast convergence by proceeding time-sharing
among the ongoing learned new policy, the DK policy and old policies.

\subsubsection{Details of the Proposed Hybrid RL Framework}

In the hybrid RL framework, there are $N\geq1$ old policies $\pi_{1},...,\pi_{N}$
learned under other similar environments, a DK policy $\pi_{N+1}$
and a new policy $\pi_{0}$. At each time $l$, the $n$-th policy
is used with probability $p_{n}$, $n\in\left[0,...,N+1\right]$.
We use $\boldsymbol{\gamma}_{n}\in\boldsymbol{\Upsilon}$ to denote
the parameters of the DNN for the $n$-th policy with $n=0,1,...,N$.
The old policies and the DK policy are fixed and the new policy together
with the reuse probabilities $\boldsymbol{p}=\left[p_{0},p_{1},...,p_{N+1}\right]^{T}$
will be updated.

Specifically, for the old policies and the new policy, we employ the
commonly used Gaussian policy \cite{sutton2018reinforcement} with
mean $\boldsymbol{\mu}_{n}$ and diagonal elements of the covariance
matrix $\mathbf{\Sigma}_{n}$ parameterized by $\boldsymbol{\mu}_{n}=f_{m_{\boldsymbol{\mu}}}\bigl(\boldsymbol{\gamma}_{n};\boldsymbol{s}\bigr)\subseteq\mathbb{R}^{n_{\boldsymbol{a}}}$
and $\text{Diag}\left(\mathbf{\Sigma}_{n}\right)=f_{m_{\boldsymbol{\sigma}}}\bigl(\boldsymbol{\gamma}_{n};\boldsymbol{s}\bigr)\subseteq\mathbb{R}^{n_{\boldsymbol{a}}}$,
respectively, and keep the non-diagonal elements of $\mathbf{\Sigma}_{n}$
as $0$. That is,
\begin{equation}
\pi_{n}\left(\boldsymbol{a}\mid\boldsymbol{s}\right)\propto\left|\mathbf{\Sigma}_{n}\right|^{\text{\ensuremath{-\frac{1}{2}}}}\mathrm{exp}\Bigl(-\frac{1}{2}\bigl(\boldsymbol{\mu}_{n}-\boldsymbol{a}\bigr)^{\intercal}\mathbf{\Sigma}_{n}^{-1}\bigl(\boldsymbol{\mu}_{n}-\boldsymbol{a}\bigr)\Bigr).\label{eq:pi}
\end{equation}

On the other hand, the DK policy is a deterministic policy in which
the action (the weight vector) is chosen based on the queue length,
i.e.,
\[
\boldsymbol{a}_{l}=\left\{ Q_{i}^{s}\left(l\right)\right\} .
\]
where $Q_{i}^{s}\left(l\right)$ denotes the queue length of user
$i$ at the $l$-th time slot, and is equal to $\sum_{l'=0}^{D_{i}-1}Q_{i,l-l'}$.
And the greedy user scheduling algorithm is then applied to maximize
the WSR based on $\boldsymbol{a}_{l}$. The Q-weighted greedy scheduling
algorithm is shown to perform well for light to moderate traffic loading,
and can provide a stable scheduling performance. To facilitate the
derivation of policy reuse probability gradient, we approximate the
deterministic Q-weighted policy using the Gaussian policy in (\ref{eq:pi})
with the mean $\boldsymbol{\mu}_{N+1}$ given by the action from the
deterministic policy, and the variance set to be a small value.

Let $\pi_{\boldsymbol{\gamma}_{0}}=\pi_{0}$ denote the new policy
with parameter $\boldsymbol{\gamma}_{0}$. Then the hybrid policy
can be expressed as
\begin{equation}
\pi_{\boldsymbol{\theta}}=\sum_{n=0}^{N}p_{n}\pi_{\boldsymbol{\gamma}_{n}}+p_{N+1}\pi_{N+1}.\label{eq:mixP-1}
\end{equation}
Note that the parameters of the old policies and the DK policy are
fixed, while the reusing probability $\boldsymbol{p}$ and the parameter
of the new policy are updated by optimizing (\ref{eq:objective}).
Thus, we omit the parameters of old polices and the DK policy, and
the hybrid policy is expressed as $\pi_{\boldsymbol{\theta}}$ with
parameters $\boldsymbol{\theta}=\left[\boldsymbol{p};\boldsymbol{\gamma}_{0}\right]\in\boldsymbol{\Theta}$.

It is a non-trivial task to solve Problem (\ref{eq:objective}) under
the hybrid policy $\pi_{\boldsymbol{\theta}}$. Although the hard
latency constraints are embodied in the objective, avoiding the use
of CMDP, the unpredictable burst traffic with multi-timeslots transmissions
also brings sparse rewards, which poses challenges for fast-convergent
algorithm design.

\subsubsection{Comparison of the Proposed Framework and the Policy Reuse Method}

Both the proposed HRL-RSHLC and the policy reuse
method can be viewed as stochastic policies, whose actions are generated
from a mixture distribution (e.g., mixture Gaussian policy) instead
of a single distribution, e.g., the Gaussian policy \cite{sutton2018reinforcement}.
However, there are significant differences between the proposed hybrid
RL framework and the policy reuse method:
\begin{itemize}
\item \textbf{The reusing probability}:
The probability of using each sub-policy, i.e., following each sub-distribution,
is determined by a heuristic method and the convergence of the overall
algorithm is not guaranteed in policy reuse method. In other words,
only the parameter of the new policy is optimized based on RL. While
the hybrid RL framework essentially can be viewed as a stochastic
policy with parameters $\boldsymbol{\theta}=\left[\boldsymbol{p};\boldsymbol{\gamma}_{0}\right]\in\boldsymbol{\Theta}$
to be optimized based on the policy gradient, where $\boldsymbol{\gamma}_{0}$
is the parameter of the new policy and $\boldsymbol{p}$ is the probability
of using each sub-policy. Unlike policy reuse method, the reusing
probability is updated along with the new policy by solving the corresponding
MDP problem, which is proved to converge to KKT points in this paper.
\item \textbf{The reusing policies}:
The policy reuse method only reuse old similar policies, while the
hybrid RL framework contains not only old similar policies but also
the DK policy to improve robustness. On one hand, the proposed framework
can utilize past learned similar policies to accelerate the convergence
when the new policy is not well studied initially, with rigorous convergence
analysis and theoretical performance guarantee. On the other hand,
it makes use of the DK policy to improve robustness to avoid any catastrophic
failure of DNN-based policies even if all the DNN-based policies cannot
work well (e.g., due to the limited representational capacity of practical
DNNs and the lack of interpretability), making it suitable for online
resource scheduling.
\end{itemize}

\section{The Proposed HRL-RSHLC Algorithm\label{sec:algorithm}}

In this section, we propose an algorithm called hybrid RL-based resource
scheduling for hard latency constraints (HRL-RSHLC) to solve the Problem
(\ref{eq:objective}). In the hybrid RL framework, there are $N\geq1$
old policies $\pi_{1},...,\pi_{N}$ trained under other similar environments
(parameterized by DNNs), a DK policy $\pi_{N+1}$ , and the new policy
$\pi_{0}\triangleq\pi_{\gamma_{0}}$ (parameterized by DNN with parameter
$\boldsymbol{\gamma}_{0}$). In each iteration $l$, the agent randomly
chooses policy $\pi_{m},m\in\left[0,1,...,N,N+1\right]$ with probability
$p_{m}$. Then the agent generates the action $\boldsymbol{a}_{l}$
according to $\pi_{m}$ based on the current state $\boldsymbol{s}_{l}$,
interacts with environment and obtains the cost $C\left(s_{l},a_{l}\right)$,
and updates the data storage $\psi_{l}$. Finally, $\psi_{l}$ is
used to update the hybrid policy $\pi_{\boldsymbol{\theta}}$. In
the following, we elaborate on the update of the policy in detail.

\subsection{Summary of HRL-RSHLC Algorithm}

We adopt the\textcolor{black}{{} stochastic successive convex approximation
(SSCA)} method to handle the stochasticity and the non-convexity of
the Problem (\ref{eq:objective}). Specifically, at each iteration,
the objective function $J\left(\boldsymbol{\theta}\right)$ are firstly
replaced by a convex surrogate function $J_{c}^{l}\left(\boldsymbol{\theta}\right)$,
which is constructed by the estimated function value $\tilde{J}^{l}$
and the estimated gradient $\tilde{\boldsymbol{g}}^{l}$. Then, we
can address the original problem (\ref{eq:objective}) by solving
a sequence of convex surrogate optimization problems. The convex surrogate
function $J_{c}^{l}\left(\boldsymbol{\theta}\right)$ can be seen
as a convex approximation of $J\left(\boldsymbol{\theta}\right)$
based on the $l$-th iterate $\boldsymbol{\theta}^{l}$, which can
be formulated as:
\begin{equation}
J_{c}^{l}\left(\boldsymbol{\theta}\right)=\tilde{J}^{l}+\left(\tilde{\boldsymbol{g}}^{l}\right)^{T}\left(\boldsymbol{\theta}-\boldsymbol{\theta}^{l}\right)+\varsigma||\boldsymbol{\theta}-\boldsymbol{\theta}^{l}||_{2}^{2},\label{surrogate-1}
\end{equation}
where $\tilde{J}^{l}\in\mathbb{R}$ and $\tilde{\boldsymbol{g}}^{l}\in\mathbb{R}^{n_{\boldsymbol{\theta}}}$
are the estimate of function value $J\left(\boldsymbol{\theta}\right)$
and the estimate of gradient $\nabla J\left(\boldsymbol{\theta}\right)$
at the $l$-th iteration, and $\varsigma$ is a positive constant.
$\tilde{J}^{l}$ and $\tilde{\boldsymbol{g}}^{l}$ are updated according
to
\begin{equation}
\tilde{J}^{l}=\chi_{l}\bar{J}^{l}+\left(1-\chi_{l}\right)\tilde{J}^{l-1},\label{estimate function-1}
\end{equation}
\begin{equation}
\tilde{\boldsymbol{g}}^{l}=\chi_{l}\bar{\boldsymbol{g}}^{l}+\left(1-\chi_{l}\right)\tilde{\boldsymbol{g}}^{l-1},\label{estimate gradient-1}
\end{equation}
where $\left\{ \chi_{l}\right\} $ is the update step size satisfying
the Assumption 1 in Section \ref{sec:converge} and $\bar{J}^{l}$
and $\bar{\boldsymbol{g}}^{l}$ are the new expressions of estimate
of function value and estimate of gradient at the $l$-th iteration,
whose specific forms will be given in Section \ref{subsec:Estimation-of--1}.

After replacing the objective function $J\left(\boldsymbol{\theta}\right)$
with the convex surrogate function $J_{c}^{l}\left(\boldsymbol{\theta}\right)$,
the optimal solution $\boldsymbol{\theta}_{c}^{l}=\left[\boldsymbol{p}_{c}^{l};\boldsymbol{\gamma}_{0}{}_{c}^{l}\right]$
is obtained by solving the following problem:
\begin{equation}
\boldsymbol{\theta}_{c}^{l}=\arg\min_{\boldsymbol{\theta}\in\boldsymbol{\Theta}}J_{c}^{l}\left(\boldsymbol{\theta}\right).\label{eq:objective1-1}
\end{equation}
Problem (\ref{eq:objective1-1}) can be viewed as a convex approximation
of the original problem, which belongs to the convex quadratic problem
and the closed-form solution can be easily obtained. Then parameter
of the new policy $\boldsymbol{\gamma}_{0}$ is updated according
to
\begin{equation}
\boldsymbol{\gamma}_{0}^{l+1}=\eta_{l}\boldsymbol{\gamma}_{0}{}_{c}^{l}+\left(1-\eta_{t}\right)\boldsymbol{\gamma}_{0}^{l},\label{update-theta-1}
\end{equation}
where $\left\{ \eta_{l}\right\} $ is the update step size satisfying
the Assumption 1 in Section \ref{sec:converge}. Note that we consider
the hybrid policy with $\boldsymbol{\theta}=\left[\boldsymbol{p};\boldsymbol{\gamma}_{0}\right]\in\boldsymbol{\Theta}$,
including the policy reuse probabilities $\boldsymbol{p}=\left[p_{0},p_{1},...,p_{N+1}\right]^{T}$satisfying
$\sum_{n=0:N+1}p_{n}=1$ and $0\leq p_{n}\leq1,\forall n$. $\boldsymbol{p}^{l+1}$
is further projected to a convex set $\Gamma=\left\{ \boldsymbol{x}=\left[x_{n}\right]\in\mathbb{R}^{N+1}:\sum_{n=0:N+1}x_{n}=1,0\leq x_{n}\leq1,\forall n\right\} $:
\begin{eqnarray}
\boldsymbol{p}^{l+1} & = & \min_{\boldsymbol{x}=[x_{0},...,x_{N+1}]}||\left(1-\eta_{l}\right)\boldsymbol{\boldsymbol{p}}^{l}+\eta_{l}\boldsymbol{p}_{c}^{l}-\boldsymbol{x}||_{2}\nonumber \\
 & s.t. & 0\leq x_{n}\leq1,\forall n\in\left[0,...,N+1\right],\nonumber \\
 &  & \sum_{n=0:N+1}x_{n}=1.\label{eq:update-p}
\end{eqnarray}

Moreover, in order to accelerate the convergence of HRL-RSHLC, we
adopt the policy reuse method in \cite{fernandez2006probabilistic}
to initialize the policy reuse probabilities $\boldsymbol{p}^{0}$,
where the policy with larger sum of rewards in several time slots
will be more likely chosen.Estimation of $\bar{J}^{l}$ and $\bar{\boldsymbol{g}}^{l}$
\label{subsec:Estimation-of--1}

It is known that the hard latency constraints are embodied in the
HLC-ET, avoiding the use of CMDP. However, it also brings sparse rewards,
that is, only when a packet is delivered successfully at one time
slot, the original data size would be included in the HLC-ET, and
the reward during many time slots may equal to 0 due to the multi-timeslots
transmission of long packets . To alleviate the issues caused by the
sparsity of reward, we adopt the idea by using both stored old experiences
and newly added data to estimate the value of $\bar{J}^{l}$ and $\bar{\boldsymbol{g}}^{l}$.
Compared to other methods such as reward shaping, self-supervised
learning and so on, old experience reusing neither needs to deign
a complicated reward shaping method, nor requires to pre-train
the agent with high computational complexity, and it is shown to be
an efficient method without increasing too much complexity. Moreover,
the proposed algorithm is well-suited to the online learning scenario
when deployed in real-world systems by reusing old experiences. Specifically,
the agent stores the latest $2L$ experiences, i.e., $\psi_{l}=\left\{ \boldsymbol{s}_{l-2L+1},\boldsymbol{a}_{l-2L+1},C\left(\boldsymbol{s}_{l-2L+1},\boldsymbol{a}_{l-2L+1}\right),...,\boldsymbol{s}_{l},\boldsymbol{a}_{l},C\left(\boldsymbol{s}_{l},\boldsymbol{a}_{l}\right)\right\} $.
At the $l$-th iteration, the agent chooses policy $\pi_{m},m\in\left[0,1,...,N,N+1\right]$
with probability $p_{m}$, and generates an action $\boldsymbol{a}_{l}$
based on state $\boldsymbol{s}_{l}$, , interacts with the environment
and obtain cost $C\left(\boldsymbol{s}_{l},\boldsymbol{a}_{l}\right)$.
Then $\left\{ \boldsymbol{s}_{l},\boldsymbol{a}_{l},C\left(\boldsymbol{s}_{l},\boldsymbol{a}_{l}\right)\right\} $
is stored in $\psi_{l}$.

We save $2L$ number of experiences because the estimation of Q-value
which needs a trajectory of experiences is involved in the estimation
of gradient $\bar{\boldsymbol{g}}^{l}$. The new expression of estimate
of function value at the $l$-th iteration $\bar{J}^{l}$ is obtained
by the sample average method based on the data storage $\varepsilon_{l}$:
\begin{equation}
\bar{J}^{l}=\frac{1}{L}\sum_{r=1}^{2L}C\left(\boldsymbol{s}_{l-2L+r},\boldsymbol{a}_{l-2L+r}\right).\label{eq:newJ}
\end{equation}

According to the theorem of policy gradient in \cite{sutton2018reinforcement},
\cite{sutton1999policy}, the gradient of $J\left(\boldsymbol{\theta}\right)$
is
\begin{equation}
\nabla J\left(\boldsymbol{\theta}\right)=\mathbb{E}_{\boldsymbol{s}\sim\mathbf{P}_{\pi_{\boldsymbol{\theta}}},\boldsymbol{a}\sim\pi_{\boldsymbol{\theta}}(\cdot\mid\boldsymbol{s})}\left[Q^{\pi_{\boldsymbol{\theta}}}\left(\boldsymbol{s},\boldsymbol{a}\right)\nabla\log\pi_{\theta}\left(\boldsymbol{a}\mid\boldsymbol{s}\right)\right],\label{gradient}
\end{equation}
where $Q^{\pi_{\boldsymbol{\theta}}}\left(\boldsymbol{s},\boldsymbol{a}\right)$
is the Q-value function, which can be formulated as:
\begin{align}
Q^{\pi_{\boldsymbol{\theta}}}\left(\boldsymbol{s},\boldsymbol{a}\right) & =\nonumber \\
\mathbb{E}_{p_{s}\sim\pi_{\boldsymbol{\theta}}}\left[\sum_{r=0}^{\infty}\left(C\left(\boldsymbol{s}_{r},\boldsymbol{a}_{r}\right)-J\left(\boldsymbol{\theta}\right)\right)\mid S_{0}=\boldsymbol{s},A_{0}=\boldsymbol{a}\right]\label{eq:Q-value}
\end{align}

The new expression of estimate of gradient at the $l$-th iteration
$\bar{\boldsymbol{g}}^{l}$ is also obtained by the sample average
method:
\begin{align}
\bar{\boldsymbol{g}}^{l}=\frac{1}{L}\sum_{r=1}^{L}\tilde{Q}^{l-2L+r}\left(\boldsymbol{s}_{l-2L+r},\boldsymbol{a}_{l-2L+r}\right)\nonumber \\
\nabla\log\pi_{\boldsymbol{\theta}^{l}}\left(\boldsymbol{a}_{l-2L+r}\mid\boldsymbol{s}_{l-2L+r}\right),\label{update-d-1-2}
\end{align}
where
\begin{align}
\tilde{Q}^{l-2L+r}\left(\boldsymbol{s}_{l-2L+r},\boldsymbol{a}_{l-2L+r}\right)=\nonumber \\
\sum_{r'=0}^{L-1}\left(C\left(\boldsymbol{s}_{l-2L+r+r'},\boldsymbol{a}_{l-2L+r+r'}\right)-\bar{J}^{l}\right)\label{update-d-1-1-1}
\end{align}
is the estimate of Q-value starting from state $\boldsymbol{s}_{l-2L+r}$
and action $\boldsymbol{a}_{l-2L+r}$, which is obtained by using
a trajectory of $L$ experiences starting from state $\boldsymbol{s}_{l-2L+r}$
and action $\boldsymbol{a}_{l-2L+r}$. Note that we can generate not
only one new experience at each iteration, but also can generate multiple
new experiences, which can help to accelerate the convergence of HRL-RSHLC,
and the number of new experiences at each iteration is referred to
as batch size in this paper.

The overall algorithm is summarized in Algorithm 1.

\begin{algorithm}
\caption{\label{HRL-RSHLC}HRL-RSHLC Algorithm}

\textbf{Input:} The decreasing sequences $\left\{ \chi_{l}\right\} $
and $\left\{ \eta_{l}\right\} $, the initial policy parameters $\boldsymbol{\theta}^{0}\in\boldsymbol{\Theta}$
and first accumulate $2L$ experiences.

\textbf{for} $l=0,1,\cdots$ \textbf{do}

\phantom{} \phantom{}1.\textbf{ }Choose policy $\pi_{l}$ from $\left\{ \pi_{m},m\in[0,...,N+1]\right\} $
with probability $P(\pi_{l}=\pi_{m})=p_{m}^{l}$. Generate the action
$\boldsymbol{a}_{l}$ according to $\pi_{l}$ based on the current
state $\boldsymbol{s}_{l}$.

\phantom{} \phantom{}2.\textbf{ Environment Interaction:}

\phantom{} \phantom{}\textbf{ }\phantom{} \phantom{}(a) Obtain
the scheduling scheme by maximizing the WSR in (\ref{WSR-1}) with
weight vector $\boldsymbol{w}_{l}=\boldsymbol{a}_{l}$.

\phantom{} \phantom{}\textbf{ }\phantom{} \phantom{}(b) Obtain
the transmitted rate of each user $R_{i}\left(\boldsymbol{s}_{l},\boldsymbol{a}_{l}\right),i\in\mathcal{K}$.

\phantom{} \phantom{}\textbf{ }\phantom{} \phantom{}(c) Obtain
cost $C\left(\boldsymbol{s}_{l},\boldsymbol{a}_{l}\right).$

\phantom{} \phantom{}\textbf{ }\phantom{} \phantom{}(d) Update
the environment status.

\phantom{} \phantom{}3.\textbf{ }Update the data storage $\psi_{l}$.

\phantom{} \phantom{}4.\textbf{ }Estimate function value and gradient
by (\ref{estimate function-1}) and (\ref{estimate gradient-1}),
respectively.

\phantom{} \phantom{}5.\textbf{ }Update the surrogate function $J_{c}^{l}\left(\boldsymbol{\theta}\right)$
via (\ref{surrogate-1}).

\phantom{} \phantom{}6.\textbf{ }Solve Problem (\ref{eq:objective1-1})
to obtain $\boldsymbol{\theta}_{c}^{l}$, and update policy parameter
$\boldsymbol{\theta}^{l+1}$ according to (\ref{update-theta-1})
and (\ref{eq:update-p}).

\textbf{end for}
\end{algorithm}

Note that the complexity of the proposed hybrid RL
framework would not increase compared to the single-policy network
such as SCAOPO and heuristic reuse probability method, since only
one policy/DNN would be chose and be updated at each iteration. In
practice, we can pre-select only a few old DNN-based policies according
to some heuristic methods such as the policy reuse method. Thus, $N$
is usually a small number, e.g., 2 or 3, then the increase in the
storage requirement is totally acceptable.

\section{Convergence and Performance Analysis\label{sec:converge}}

In this section, we first present the key assumptions for convergence
analysis. Based on this, we prove that the proposed HRL-RSHLC algorithm
converges to a KKT point of the Problem (\ref{eq:objective}).

\subsection{Key Assumptions on the Problem Structure and Algorithm Parameters}

\subsubsection{Assumptions of Problem (\ref{eq:objective})}

\noindent \newtheorem{assumption}{Assumption}
\begin{assumption}\textit{ (Assumptions on the Problem Structure:)}\\
1) There exist constants $\lambda>0$ and $\rho\in\left(0,1\right)$
satisfying
\begin{equation}
\mathrm{sup}_{\boldsymbol{s}\in\mathcal{S}}d_{TV}\left(\mathbf{P}\left(\boldsymbol{s}_{l}\mid\boldsymbol{s}_{0}=\boldsymbol{s}\right),\mathbf{P_{\pi_{\boldsymbol{\theta}}}}\right)\leq\lambda\rho^{l},
\end{equation}
for all $l=0,1,\cdots$, where $\mathbf{P_{\pi_{\boldsymbol{\theta}}}}$
is the stationary state distribution under policy $\pi_{\boldsymbol{\theta}}$
and $d_{TV}\left(a,b\right)=\int_{\boldsymbol{s}\in\mathcal{S}}\left|a\left(\mathrm{d}\boldsymbol{s}\right)-b\left(\mathrm{d}\boldsymbol{s}\right)\right|$
denotes the total-variation distance between the probability measures
$a$ and $b$.\\
2) The DNNs' parameter spaces $\mathbf{\boldsymbol{\Theta}}\subseteq\mathbb{R}^{n_{\boldsymbol{\theta}}}$
for some positive integer $n_{\boldsymbol{\theta}}$, are compact
and convex, and the outputs of DNNs are bounded.\\
3) The cost/reward $C$, the derivative and the second-order derivative
of $J\left(\boldsymbol{\theta}\right)$ are uniformly bounded.

\noindent 4) The policy $\pi_{\boldsymbol{\theta}}$ follows Lipschitz
continuity over the parameter $\boldsymbol{\theta}\in\mathbf{\boldsymbol{\Theta}}$.

\noindent 5) State space $\mathcal{S}\subseteq\mathbb{R}^{n_{\boldsymbol{s}}}$
and action space $\mathcal{A}\subseteq\mathbb{R}^{n_{\boldsymbol{a}}}$
are both compact sets for some positive integers $n_{\boldsymbol{s}}$
and $n_{\boldsymbol{a}}$.\end{assumption}

Assumption 1-1) indicates that there exists the stationary state distribution
under policy $\pi_{\boldsymbol{\theta}}$, which is independent of
$\boldsymbol{s}_{0}$. And this assumption is a standard ergodicity
assumption when considering problems without episode boundaries, see
e.g.,\textcolor{magenta}{{} }\cite{zhao2019reinforcement,sutton2018reinforcement}
and \cite{zou2019finite}. Assumption 1-2) is trivial in general RL
problems. Assumption 1-3) and 1-4) indicate that the Lipschitz continuity
of $J\left(\boldsymbol{\theta}\right)$ over parameters $\boldsymbol{\theta}$,
which are always assumed in the rigorous convergence analysis of RL
algorithms \cite{zou2019finite,perkins2002convergent}, and the gradient
of the policy DNN is always finite, which can also be easily satisfied.
Assumption 1-5) considers a common scenario where the state and action
spaces can be continuous.

\subsubsection{Assumptions on the Step Sizes}

\begin{assumption}\textit{ (Assumptions on} \emph{step size}\textit{:)}\\
1) $\chi_{l}\rightarrow0$, $\frac{1}{\chi_{l}}\leq O\left(l^{\kappa}\right)$
for some $\kappa\in\left(0,1\right)$, $\sum_{l}\chi_{l}l^{-1}<\infty$,
$\sum_{l}\left(\chi_{l}\right)^{2}<\infty$, $\sum_{l}\chi_{l}(\log^{2}l\sum_{l'=l-\log l}^{l}\eta_{l'})<\infty$.\\
2) $\eta_{l}\rightarrow0$, $\sum_{l}\eta_{l}=\infty$, $\sum_{l}\left(\eta_{l}\right)^{2}<\infty$.\\
3) $\lim_{l\rightarrow\infty}\frac{\eta_{l}}{\chi_{l}}=0.$

\end{assumption}

Note that $\frac{1}{\chi_{l}}\leq O\left(l^{\kappa}\right)$ for some
$\kappa\in\left(0,1\right)$ in Assumption 2-1) is almost the same
as $\sum_{l}\chi_{l}=\infty$, which is a common assumption in stochastic
optimization algorithms \cite{liu2019stochastic}. A typical choice
of $\left\{ \chi_{l}\right\} $ and $\left\{ \eta_{l}\right\} $ satisfying
Assumption 2 is $\chi_{l}=l^{-\kappa_{1}}$ and $\eta_{l}=l^{-\kappa_{2}}$,
where $\kappa_{1}\in\left(0.5,1\right),\kappa_{2}\in(0.5,1]$ and
$\kappa_{1}<\kappa_{2}$.

\subsection{Convergence of the HRL-RSHLC}

Based on Assumptions 1 and 2, we can first prove a lemma which indicates
the asymptotic consistency of the surrogate function value $\tilde{J}^{l}$
and the surrogate gradient $\tilde{\boldsymbol{g}}^{l}$.

\noindent \newtheorem{lemma}{Lemma}

\begin{lemma} \textit{(Asymptotic consistency of surrogate functions:)}\\
Suppose that Assumptions 1 and 2 are satisfied, we have
\begin{align}
\underset{l\rightarrow\infty}{\mathrm{lim}} & \left|J\left(\boldsymbol{\theta}_{l}\right)-\tilde{J}^{l}\right|=0\label{eq:JJ}\\
\underset{l\rightarrow\infty}{\mathrm{lim}} & \Bigl\Vert\nabla J\left(\boldsymbol{\theta}_{l}\right)-\tilde{\boldsymbol{g}}^{l}\Bigr\Vert_{2}=0\label{eq:gg}
\end{align}
\end{lemma}

Please refer to Section I of our supplementary material for the proof.
Then, we consider a subsequence $\left\{ \boldsymbol{\theta}^{l_{j}}\right\} _{j=1}^{\infty}$,
which converges to a limiting point $\boldsymbol{\theta}^{*}$ when
$j\rightarrow\infty$. There exist a converged surrogate function
$\tilde{J}\left(\boldsymbol{\theta}\right)$ such that
\begin{equation}
\underset{j\rightarrow\infty}{\mathrm{lim}}J_{c}^{l_{j}}\left(\boldsymbol{\theta}\right)=\tilde{J}\left(\boldsymbol{\theta}\right),\forall\boldsymbol{\theta}\in\boldsymbol{\Theta},
\end{equation}
where
\[
\left|J\left(\boldsymbol{\theta}^{*}\right)-\tilde{J}\left(\boldsymbol{\theta}^{*}\right)\right|=0,
\]
\[
\left|\left|\nabla J\left(\boldsymbol{\theta}^{*}\right)-\nabla\tilde{J}\left(\boldsymbol{\theta}^{*}\right)\right|\right|_{2}=0.
\]

Then, based on Assumptions 1 and 2, and the Lemma 1, we are ready
to prove the main convergence theorem, which states that Algorithm
1 is able to converge to a KKT point of Problem (\ref{eq:objective})
with an arbitrary initial point.

\noindent \newtheorem{theorem}{Theorem}

\begin{theorem} \textit{(Global Convergence of Algorithm 1:) }\\
Suppose Assumptions 1 and 2 are satisfied, and since problem (\ref{eq:objective})
is an MDP without constraints, an arbitrary initial point $\boldsymbol{\theta}^{0}$
is feasible. Denote $\left\{ \boldsymbol{\theta}^{l}\right\} _{l=1}^{\infty}$
as the sequence generated by Algorithm 1 with a adequately small $\eta_{0}$.
We denote $L_{l}$ as the number of data samples, which is set to
$O\left(\textrm{log}l\right)$. Then every limiting point $\boldsymbol{\theta}^{*}$
of $\left\{ \boldsymbol{\theta}^{l}\right\} _{l=1}^{\infty}$ is a
KKT point of Problem (\ref{eq:objective}) almost surely\textcolor{blue}{.}\end{theorem}

Please refer to Section II of our supplementary material for the proof.
Note that in order to achieve the rigorous convergence proof, we assume
that $\eta_{0}$ is adequately small. Although theoretically $L_{l}$
approaches infinity when $l\rightarrow\infty$, the increasing rate
is on the logarithmic order and is relatively slow. In practice, we
notice that Algorithm 1 can still achieve a good convergence behavior
when $L_{l}$ is set to a constant number.
\begin{rem}
\textit{(Key difference from the related work) }Our previous work
\cite{tian2022successive} is closely related to the work in this
paper. However, the proposed algorithm in \cite{tian2022successive}
only optimizes the new policy without introducing DK policies and
other old policies, while the HRL-RSHLC optimizes both the new policy
and the probabilities of reusing the old policies. The innovative
design of HRL-RSHLC helps to achieve superior performance with faster
convergence speed and lower interaction cost.
\end{rem}

\section{Simulation and Performance Evaluation\label{sec:Simulation-Results}}

\subsection{Simulation Setup\label{subsec:Simulation-Setup}}

In the simulations, we adopt the clustered delay line B (CDL-B) model
\cite{3gpp.38.901} to generate the channel of users. For each user
$i$, data packets whose lengths follow a Poisson arrival distribution
with mean $\lambda_{i}$ arrive at the start of each timeslot with
probability $PA$. We set the constant in the surrogate problem as
$\zeta=1$, and choose the step sizes as $\chi_{l}=\frac{1}{l^{0.6}}$,
$\eta_{l}=\frac{1}{l^{0.7}}$. We set $K=\{8,10\}$ and $N_{t}=\{4,5\}$,
and the simulation parameters are shown in Table \ref{tab:parameters}.
In simulations, we have chosen various values of $\{D_{i}\}$, $\{\lambda_{i}\}$
and $PA$ according to the parameters reported in \cite{li2023embb},
which are typical values in burst traffic application scenarios, and
the proposed algorithm works under all of those traffic conditions.
We report in this section, the simulation results under two configurations:
(1) $\{D_{i}\}=\{4,5,6,7,4,5,6,7\}$ timeslots,$\{\lambda_{i}\}=\{22,42,62,82,22,42,62,82\}$
Kbit, $PA=0.3$; (2) $\{D_{i}\}=\{4,5,6,7,4,5,6,7\}$ timeslots,$\{\lambda_{i}\}=\{30,50,70,90,30,50,70,90\}$
Kbit, $PA=0.3$.

\begin{table}[t]
\centering{}\caption{\textcolor{blue}{\label{tab:parameters}}Simulation Parameters}
\begin{tabular}{|c|c|}
\hline 
Parameter & Value\tabularnewline
\hline 
\hline 
{\scriptsize Bandwidth} & {\scriptsize 58 MHz}\tabularnewline
\hline 
{\scriptsize TX power} & {\scriptsize 12 dBm}\tabularnewline
\hline 
{\scriptsize Antenna array} & {\scriptsize ULA}\tabularnewline
\hline 
{\scriptsize Carrier frequency} & {\scriptsize 3GHz}\tabularnewline
\hline 
{\scriptsize Path loss} & {\scriptsize{[}130,150{]} dB}\tabularnewline
\hline 
{\scriptsize$\tau$} & {\scriptsize 1ms}\tabularnewline
\hline 
{\scriptsize Number of users} & {\scriptsize 8,10}\tabularnewline
\hline 
{\scriptsize Distribution of packets} & {\scriptsize Poisson distribution}\tabularnewline
\hline 
\end{tabular}
\end{table}

\subsection{Reference Baselines}

We choose several state-of-the-art RL-based resource allocation methodologies
as baselines, to demonstrate the novelty and competitiveness of the
proposed algorithm: Soft Actor-Critic (SAC) \cite{tang2021novel}
is a state-of-art DRL algorithm that solves both discrete and continuous
control problems, and uses a stochastic policy and has been widely
used in resource allocation problems; SCAOPO \cite{tian2022successive}
is based on a single policy and adopts the constrained stochastic
successive convex approximation (CSSCA) method \cite{liu2019stochastic}
to handle the stochasticity and the non-convexity of MDP problem;
Policy reuse method has a similar framework with HRL-RSHLC, which
is a stochastic policy with mixed sub-distributions, but its reusing
probability is updated in a heuristic way according to the principle
given in \cite{fernandez2006probabilistic}; PPO is a popular RL algorithm
due to its simplicity and effectiveness in handling continuous control
tasks in complex environments. We choose Q-weighted greedy algorithm
in \cite{venkatraman2014low} as a baseline as well as the DK policy
for reuse, which is shown to perform well for light to moderate traffic
loading and can provide a stable scheduling performance. We summarized
the key differences of those RL algorithms in Table \ref{tab:Key differences},
and the unique feature of the proposed algorithm is that it is the
only RL algorithm which reuses old policies to accelerate convergence
as well as has theoretical performance guarantee.
\begin{table}[t]
\centering{}\caption{\textcolor{blue}{\label{tab:Key differences}}Key Differences of RL
Algorithms}
\begin{tabular}{|c|c|c|}
\hline 
{\scriptsize Algorithm} & {\scriptsize Reuse policies} & {\scriptsize Performance guarantee}\tabularnewline
\hline 
\hline 
{\scriptsize HRL-RSHLC} & $\checked$ & $\checked$\tabularnewline
\hline 
{\scriptsize SCAOPO} & {\scriptsize no} & $\checked$\tabularnewline
\hline 
{\scriptsize Policy reuse method} & $\checked$ & {\scriptsize no}\tabularnewline
\hline 
{\scriptsize SAC} & {\scriptsize no} & $\checked$\tabularnewline
\hline 
{\scriptsize PPO} & {\scriptsize no} & $\checked$\tabularnewline
\hline 
\end{tabular}
\end{table}
\begin{figure}
\centering
\subfloat[\label{fig:a-1}Comparison of packet loss probability.]{\centering{}\includegraphics[width=6cm,totalheight=5.5cm,height=5.2cm]{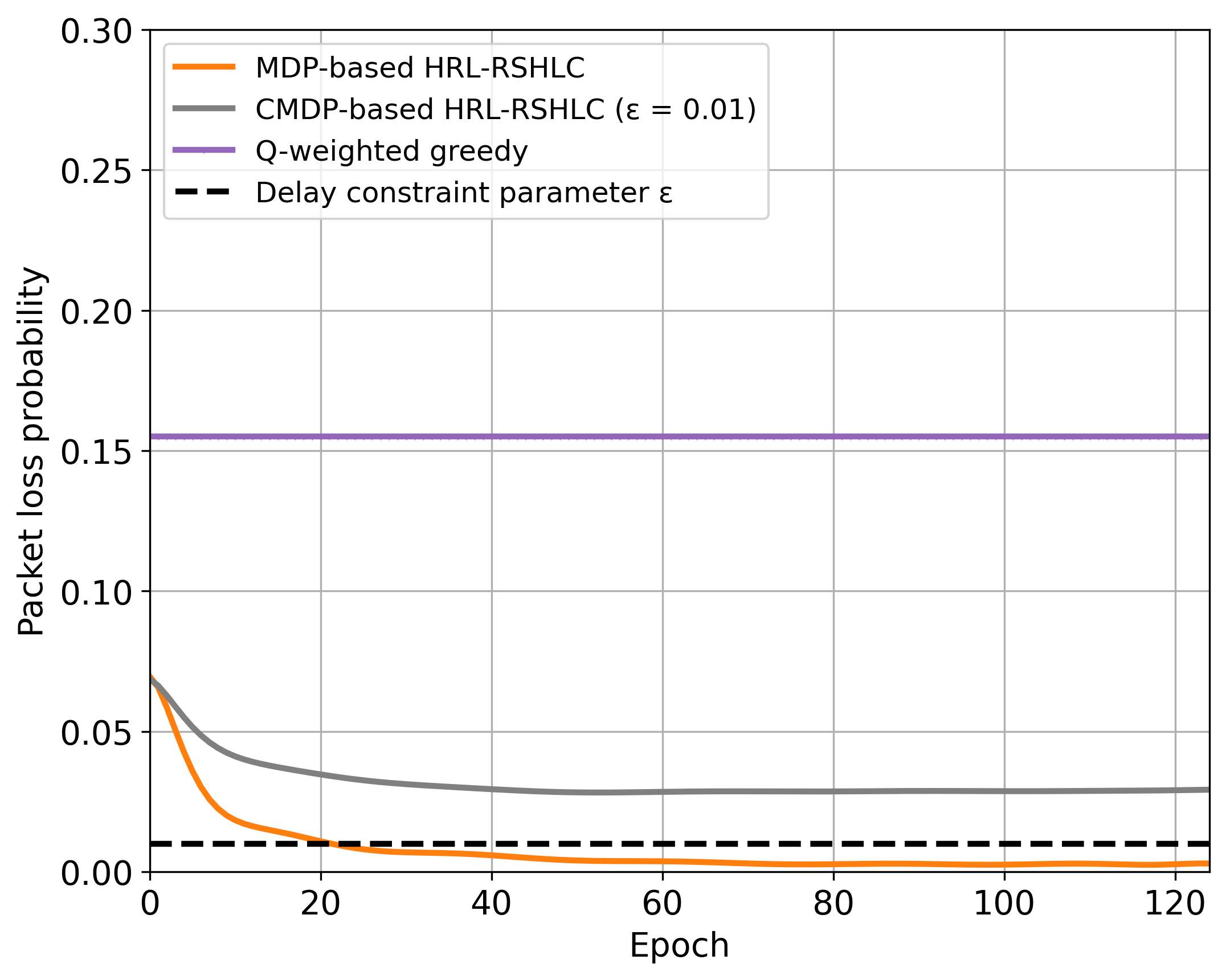}\textcolor{blue}{}}

\subfloat[\label{fig:b-1}Comparison of hard-latency constrained effective throughput.]{\begin{centering}
\includegraphics[width=6cm,totalheight=5.5cm,height=5.2cm]{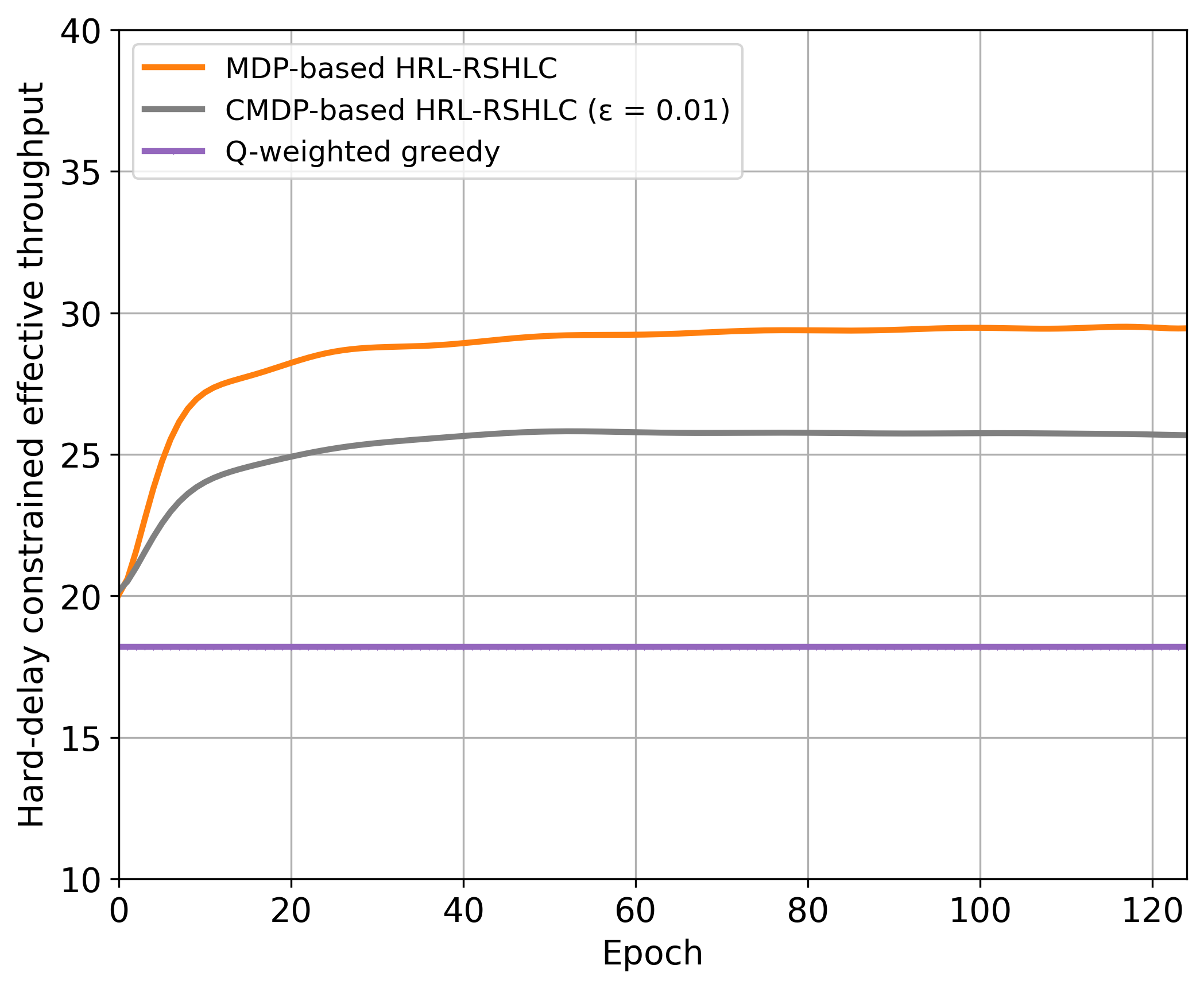}
\par\end{centering}
\textcolor{blue}{}}

\caption{\label{fig:simulation results-1}\textcolor{blue}{{} }Comparison with
CMDP-based HRL-RSHLC.}
\end{figure}
\begin{figure}[tp]
\begin{centering}
\textsf{\includegraphics[width=6cm,totalheight=5.5cm,height=5.2cm]{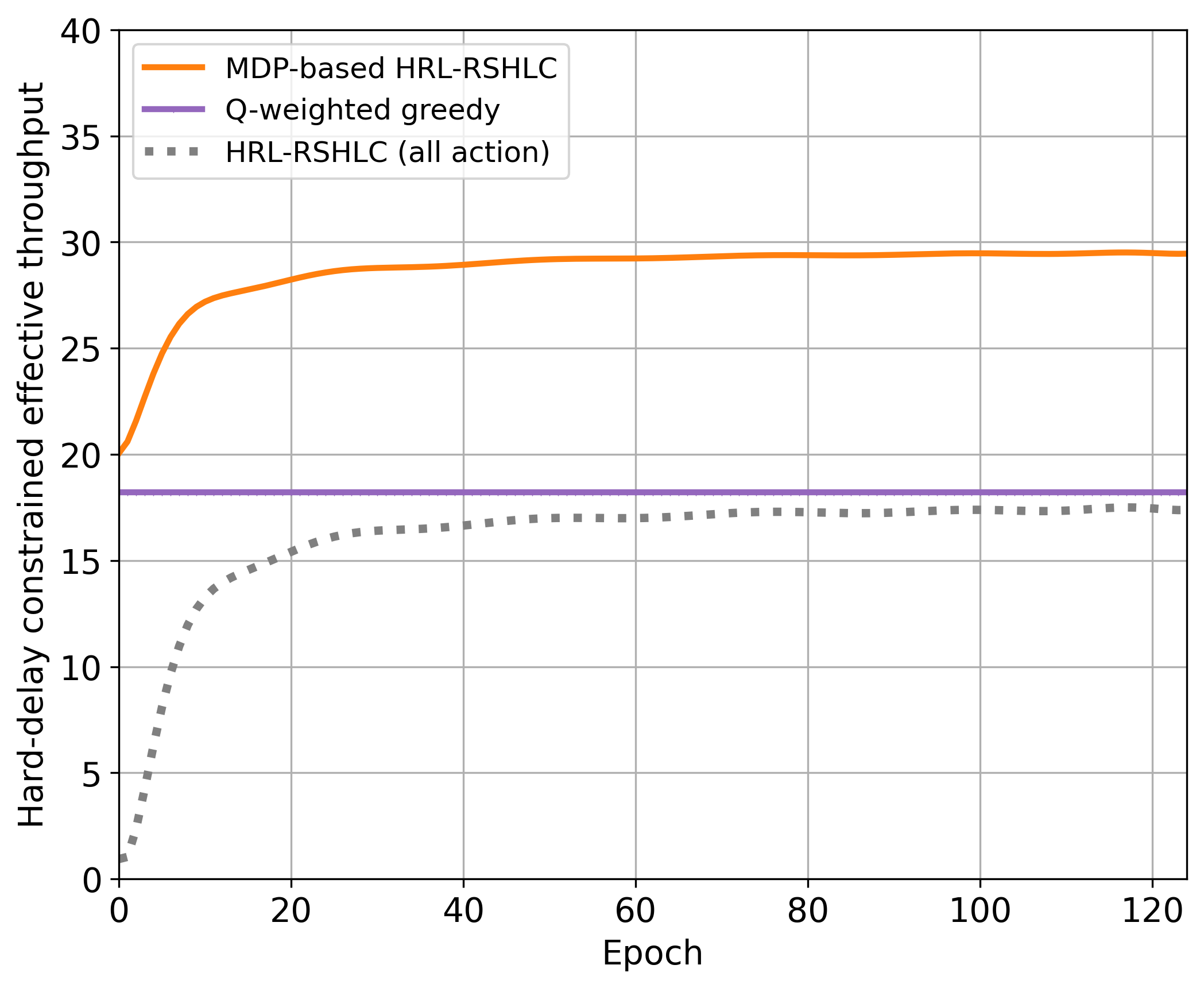}}
\par\end{centering}
\caption{\label{fig:action comparison} Comparison with HRL-RSHLC that directly
controls the user scheduling and power allocation.}
\end{figure}

\subsection{Simulation Results and Discussions}

First, the comparisons with CMDP-based HRL-RSHLC and HRL-RSHLC which
directly controls the user scheduling and power allocation are presented,
to illustrate the motivation of using MDP and choosing the weight
as the action. Then, we compare the performance of the proposed algorithm
with other baselines under different traffic conditions. Finally,
we conduct ablation experiments to better demonstrate the significance
of introducing the DK policy and old policies.

\subsubsection{Comparisons with CMDP-based HRL-RSHLC}

Fig. \ref{fig:a-1} and Fig. \ref{fig:b-1} compare packet loss probability
and hard-latency constrained effective throughput, respectively. CMDP-based
HRL-RSHLC adopts the average throughput as the reward, and the parameter
$\epsilon$ is set to 0.01. Simulation results show that CMDP-based
HRL-RSHLC has not fully converged within 130 epochs and has not achieved
the required packet loss probability, while MDP-based HRL-RSHLC which
aims to maximize the average effective throughput without an explicit
constraint has achieved lower packet loss probability as well as higher
effective throughput.

\subsubsection{Comparisons with Other Actions}

Fig. \ref{fig:action comparison} compares the learning curve of HRL-RSHLC
that only controls the priority weight with HRL-RSHLC that directly
controls the user scheduling and power allocation, and the black stars
mark the points of convergence of RL algorithms (the RL algorithm
is considered converged when the reward fluctuates around a relatively
stable level). It can be seen that after 130 epoch, HRL-RSHLC controlling
all actions has not converged to a good solution and even performs
worse than the Q-weighted greedy, which is probably due to that its
enlarged action space makes it easier to get stuck in a bad local
optimum. Those simulation results show that reducing the action space
via only controlling the priority weight can significantly improve
the convergence speed and the performance after convergence.

\subsubsection{Comparisons with Baselines}

\begin{figure}
\centering
\subfloat[\label{fig:a}The learning curve of hard-latency constrained effective
throughput under configuration (1).]{\centering{}\includegraphics[width=6cm,totalheight=5.5cm,height=5.2cm]{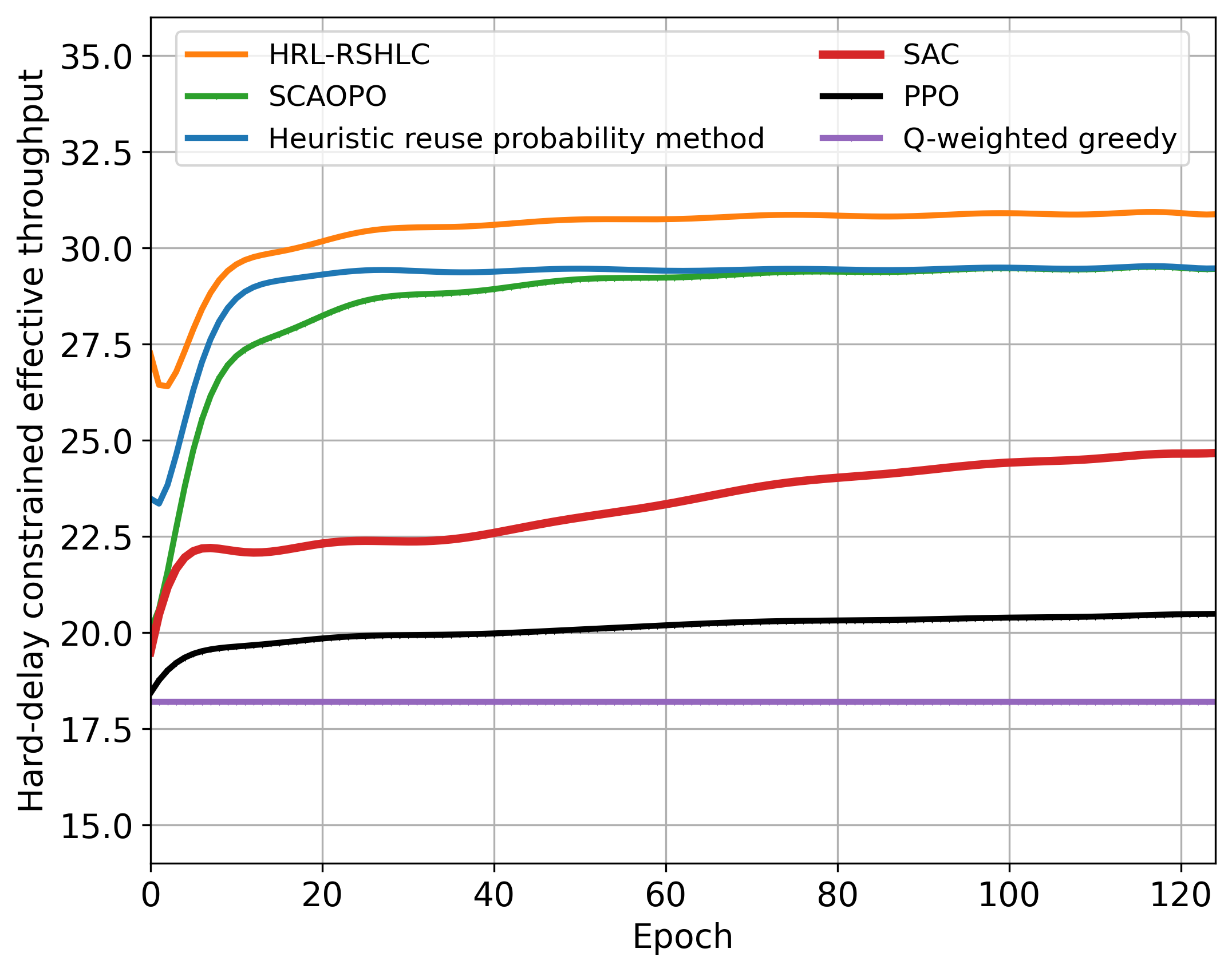}}

\subfloat[\label{fig:b}The learning curve of hard-latency constrained effective
throughput under configuration (2).]{\begin{centering}
\includegraphics[width=6cm,totalheight=5.5cm,height=5.2cm]{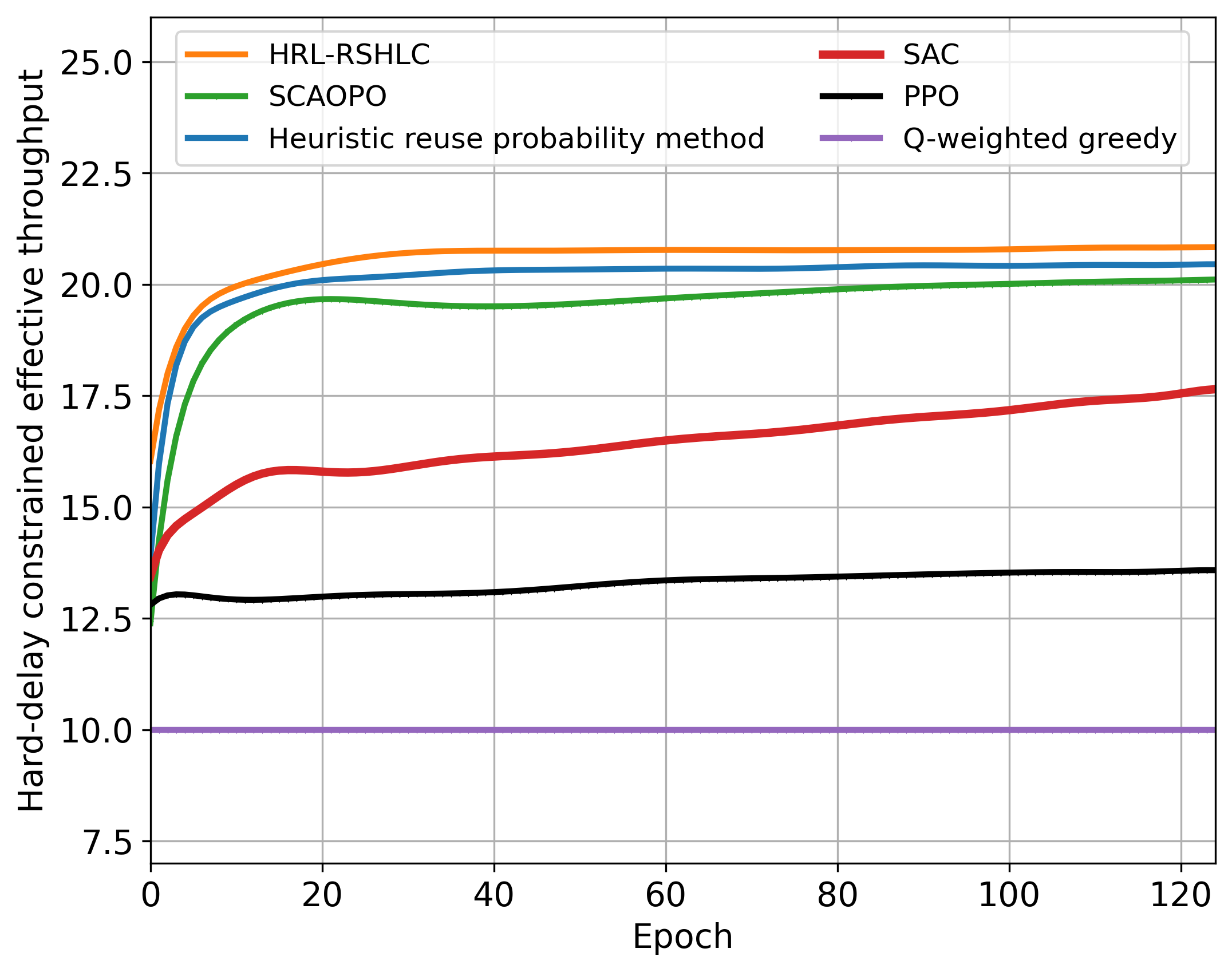}
\par\end{centering}
}

\caption{\label{fig:simulation results}Performance comparisons for different
cases when $K=8$, $N_{t}=4$.}
\end{figure}

We run the experiment under two traffic conditions given in Section
\ref{subsec:Simulation-Setup} with different user numbers and antenna
numbers. Fig. \ref{fig:a} and Fig. \ref{fig:b} show the learning
curves of hard-latency constrained effective throughput under different
traffic conditions when $K=8$, $N_{t}=4$. The proposed HRL-RSHLC
converges fastest to the highest value compared to other baselines.
SCAOPO converges faster than SAC and PPO, but slower than HRL-RSHLC
and heuristic reuse probability method. Note that although heuristic
reuse probability method achieves the second highest effective throughput
under Configuration (2), it has no rigorous convergence analysis or
theoretical performance guarantee for general cases.

\begin{figure}
\centering
\subfloat[\label{fig:a-2} The learning curve of hard-latency constrained effective
throughput under configuration (1).]{\centering{}\textcolor{blue}{\includegraphics[width=6cm,totalheight=5.5cm,height=5.2cm]{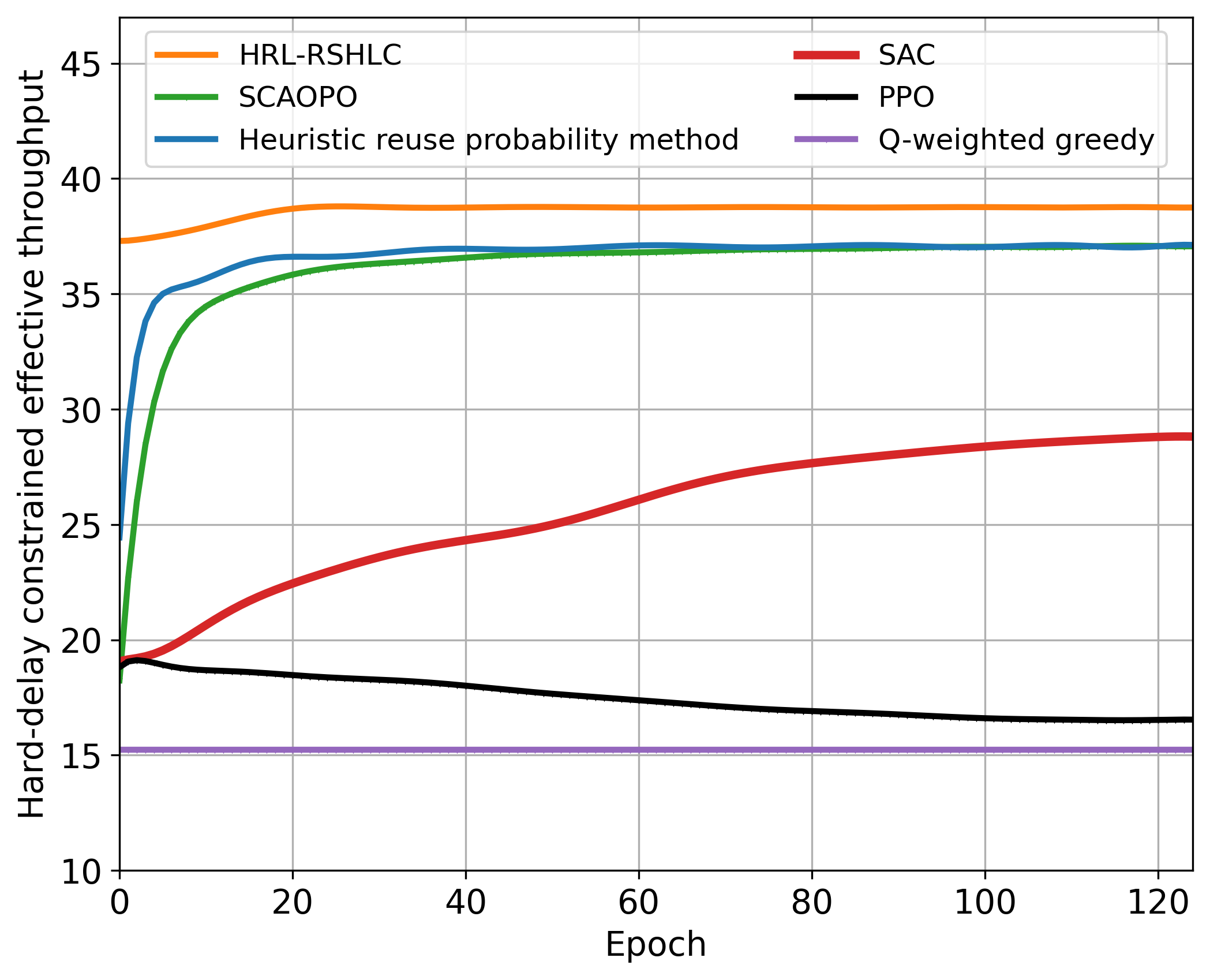}}}

\subfloat[\label{fig:b-2}The learning curve of hard-latency constrained effective
throughput under configuration (2).]{\begin{centering}
{\includegraphics[width=6cm,totalheight=5.5cm,height=5.2cm]{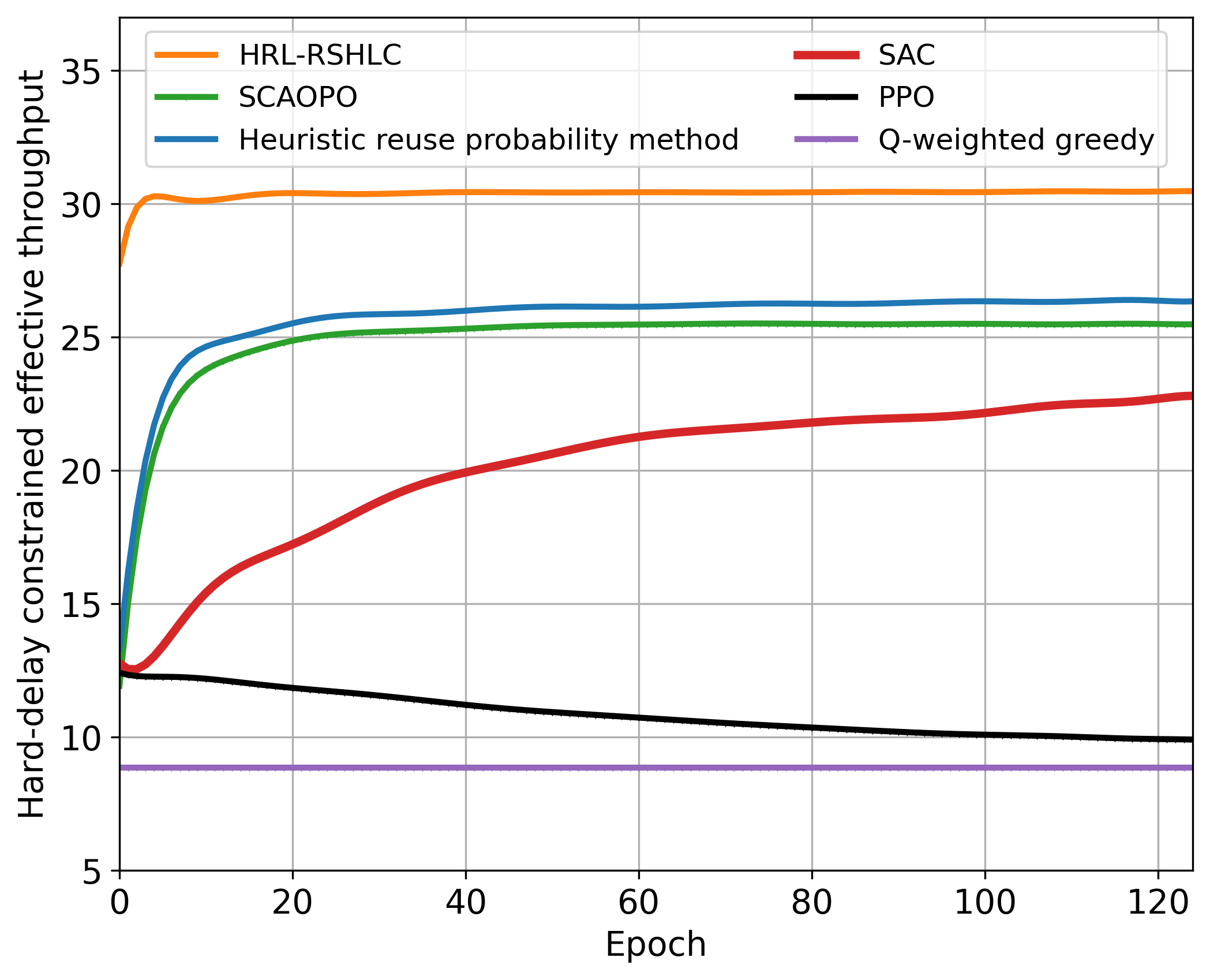}}
\par\end{centering}
\textcolor{blue}{}}

{\caption{\label{fig:simulation1}Performance comparisons for different cases
when $K=10$, $N_{t}=5$.}
}
\end{figure}

Fig. \ref{fig:a-2} and Fig. \ref{fig:b-2} show
the learning curves of hard-latency constrained effective throughput
under different traffic conditions when $K=10$, $N_{t}=5$. The proposed
HRL-RSHLC converges fastest to the highest value compared to other
baselines under both of configurations. SCAOPO converges faster than
SAC, but slower than HRL-RSHLC and heuristic reuse probability method.
PPO performs bad under this case, which may due to the increased state
and action space.

\subsubsection{Ablation Experiments}

\begin{figure}[tp]
\begin{centering}
\textsf{\includegraphics[width=6cm,totalheight=5.5cm,height=5.2cm]{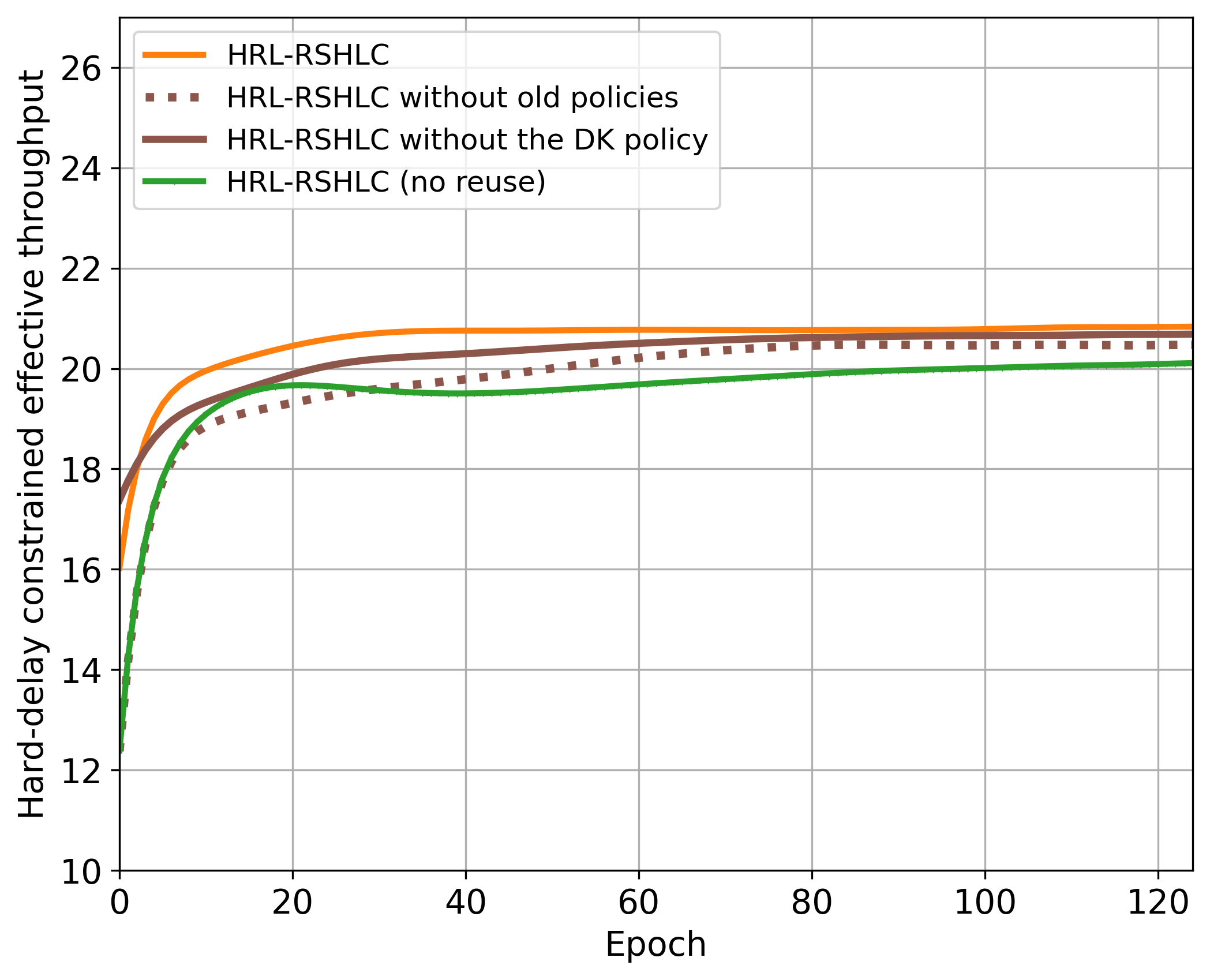}}
\par\end{centering}
\caption{\label{fig:wo} Comparison with HRL-RSHLC without old policies and
HRL-RSHLC without the DK policy.}
\end{figure}

We conduct ablation experiments to better illustrate the significance
of the DK policy and old policies. Fig. \ref{fig:wo} compares the
learning curve of proposed algorithm with that of HRL-RSHLC without
reusing old policies, HRL-RSHLC without reusing the DK policy and
HRL-RSHLC without reusing any policies. The black stars mark the points
of convergence. It is clear to see that the proposed HRL-RSHLC converges
to the best KKT point at a faster speed with the guidance of both
old policies and the DK policy, which demonstrate the benefits of
introducing both DK and old policies.

\subsubsection{Performance with imperfect CSI}

\begin{figure}[tp]
\begin{centering}
\textsf{\textcolor{blue}{\includegraphics[width=6cm,totalheight=7cm,height=4.7cm]{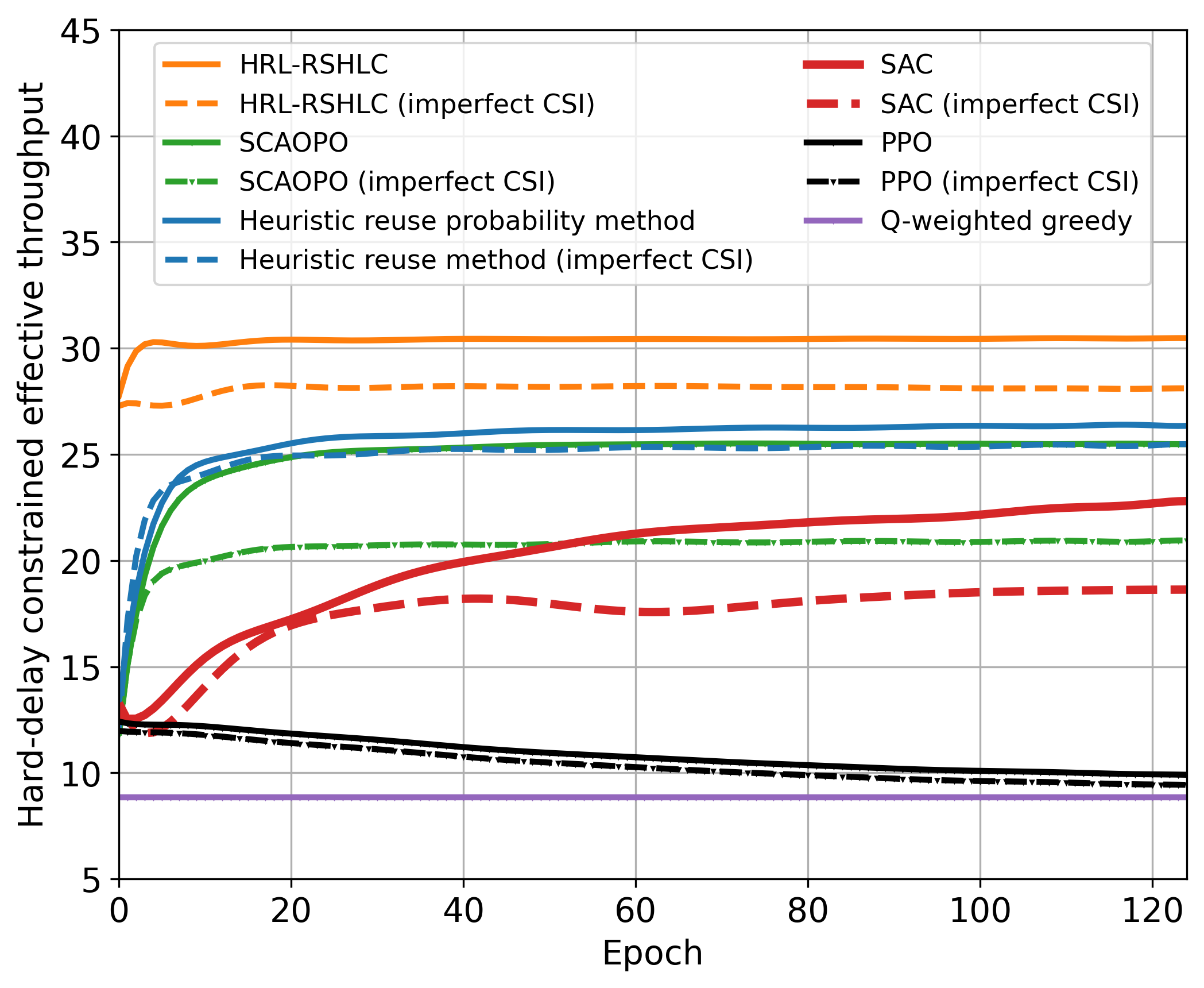}}}
\par\end{centering}
{\caption{\label{fig:simulation3}Performance comparison with imperfect CSI.}
}
\end{figure}

In the above simulations we assume perfect CSI since
the channel estimation process is independent with the resource scheduling
process so we focus on the performance of the proposed scheduling
algorithm without the added complexity of imperfect CSI. However,
to demonstrate the robustness of the proposed algorithm, we further
compare the performance of the algorithms with imperfect CSI. Specifically,
we assume the BS performs the scheduling algorithm based on imperfect
estimated CSI $\hat{\boldsymbol{H}}$, which can be expressed as

\begin{equation}
\hat{\boldsymbol{H}}=\boldsymbol{H}+\boldsymbol{n}_{e}
\end{equation}
where $\boldsymbol{n}_{e}$ is the estimation noise. In the simulations,
we adopted the Guassion noise
to act as the channel estimation error, with normalized mean square
error $\textrm{NMSE}=\frac{\mathbb{E}\left[||\boldsymbol{H}-\hat{\boldsymbol{H}}||^{2}\right]}{\mathbb{E}\left[||\boldsymbol{H}||^{2}\right]}$
of 0.4. Fig. \ref{fig:simulation3} shows the performance comparison
of the algorithms with perfect CSI. The result demonstrates that although
all DNN-based algorithms perform worse with imperfect CSI, the proposed
HRL-RSHLC still converges faster and better than other baselines.


\section{Conclusions\label{sec:conclusion}}

We propose a novel HRL-RSHLC algorithm for resource scheduling with
hard latency constraints, which reuses polices from both old policies
and domain-knowledge-based policies to improve the performance. The
joint optimization of the policy reuse probabilities and new policy
is formulated as an \textcolor{black}{MDP,} which maximizes the hard-latency
constrained effective throughput (HLC-ET) of users. In particular,
the hard delay constraints are embodied in the objective function,
which avoid the use of CMDP. SSCA\textcolor{black}{{} is applied to
handle the non-convex stochastic }characteristic of the MDP. We prove
that the proposed HRL-RSHLC can converge to KKT points with an arbitrary
initial point. Simulations show that HRL-RSHLC can achieve superior
performance with faster convergence speed. However,
the performance of the proposed algorithm may degrade in the non-stationary
case, which is also true for most resource allocation algorithms.
Future research may adopt the efficient context-aware meta-learning
to address this issue.

\bibliographystyle{IEEEtran}
\bibliography{RLreferences}

\end{document}